\documentclass[aps,pre,notitlepage,10pt]{revtex4-1}%
\usepackage{amssymb}
\usepackage{amsfonts}
\usepackage{amsmath}
\usepackage{xcolor}
\usepackage{graphicx}%
\providecommand{\U}[1]{\protect\rule{.1in}{.1in}}
\begin{document}
\title{Approximate calculation of multidimensional first passage times}
\author{James F. Lutsko}
\homepage{http://www.lutsko.com}
\affiliation{Center for Nonlinear Phenomena and Complex Systems CP 231 and BLU-ULB Space
Research Center, Universit\'{e} Libre de Bruxelles, Blvd. du Triomphe, 1050
Brussels, Belgium}
\email{jlutsko@ulb.be}

\begin{abstract}
The general, multidimensional barrier crossing problem for diffusive processes under the action of conservative forces is studied with the goal of developing tractable approximations. Particular attention is given to the effect of different statistical interpretations of the stochastic differential equation and to the relation between the approximations and the known, exact solutions to the one-dimensional problem. Beginning with a reasonable, but heuristic, simplifying assumption, a one-dimensional solution to the problem is developed. This is then simplified by introducing further approximations resulting in a sequence of increasingly simple expressions culminating in the classic result of Langer(Ann. Phys. 54, 258 (1969)) and others. The various approximations are tested on two dimensional problems by comparison to simulation results and it is found that the one-dimensional approximations capture most of the non-Arrhenius dependence on the energy barrier which is lost in the Langer approximation while still converging to the latter in the large-barrier limit.
\end{abstract}
\date{\today }
\maketitle

\section{Introduction}

Stochastic barrier crossing is a fundamental paradigm in physics that
abstracts a wide range of physical processes such as nucleation of first order
phase transitions, protein folding, chemical reactions\cite{Hanggi, Gardiner}.
Stochastic models can be developed systematically using tools from statistical
physics such as projection operators\cite{Grabert,Zubarev,Duran-Olivencia} and
multiscale analysis\cite{Jarzynski,Bendahmane}, or constructed using a
combination of physical insight about the specific processes and heuristics
such as the fluctuation-dissipation relation (e.g. Landau's introduction of
fluctuating hydrodynamics\cite{Landau}). The mean first passage time (MFPT)
for barrier crossing, or equivalently the escape rate or transition rate, is
often the most important physical quantity for practical applications. For
models involving a single stochastic variable, exact expressions for these
quantities are easily derived as discussed below. Some exact results also exist for sufficiently symmetric
multi-dimensional processes\cite{Metzler1} and as well as some general asymptotic results\cite{Metzler2}. However, the
general multidimensional case remains challenging. Indeed, aside from direct or
indirect simulation of the barrier crossing process, the most widely used
approach to extract rates from multidimensional models is the classic theory
of Langer\cite{Langer1,Langer2} which yields an approximate formula valid in
conditions corresponding (roughly) to a large minimal energy barrier and
energy surfaces that can be approximated as quadratic functions in the
proximity of the critical point. Exact

The goal of this paper is to explore the domain between the exact one
dimensional results and the classic multi-dimensional approximation. In many
physical problems it is natural to imagine that there is a well-defined
quasi-one dimensional pathway that dominates the transition process. The
scalar variable characterizing position along this pathway goes by various
names such as the reaction coordinate or the order parameter and it is often
assumed that the process can be essentially understood as a one-dimensional
problem in terms of this variable. This intuition has been formalized in
recent years particularly in the work of Berezhkovskii and
Szabo\cite{Szabo1,Szabo2} in which the order parameter is identified as the
commitor function, which is the probability that a trajectory starting at a
given point will reach one basin of attraction before another. Berezhkovskii
and Szabo show that this leads to an exact mapping of the multidimensional
problem to a one-dimensional equation, at the expense of having to know the
solution of the original problem to define the order parameter. They go on to
show that natural assumptions of a separation of timescales between the order
parameter an all other variables leads to the classic multi-dimensional result.

In the present work, similar ideas are used but with a somewhat different perspective. The goal here is to analyze the general problem and to develop a sequence of approximate solutions, each simpler and less precise than the previous one, culminating in the classic result of Langer. Having done so, the various levels of approximation are then
tested on
various two-dimensional barrier-crossing problems and compared to ``exact'' simulations. The next Section formulates the
problem in the language of the mean first passage time (MFPT) and recalls the
relation between the MFPT and the Fokker-Planck equation. Everything is
developed for the general case of state-dependent noise and the complications
connected to this (e.g. the statistical interpretation) are discussed. Section
\ref{secMFPT} continues the theoretical development first by briefly reviewing
the exactly solvable one-dimensional problem and then with the main
approximations explored here. Section \ref{secNumerical} presents the results
of some numerical experiments in two dimensions in which the various levels of
approximations are compared to simulations. For the examples discussed here,
it is found that the most general one-dimensional approximation gives a
surprisingly good estimate of the MFPT. The final Section summarizes the
conclusions of this work.

\section{Formal development}

\label{secFormal}

\subsection{Diffusive stochastic model}

Our starting point is a conservative diffusive stochastic dynamics of the
general form%
\begin{equation}
\frac{d}{dt}\widehat{x}^{I}=g^{IJ}\left(  \widehat{x}\right)  \left.
\frac{\partial}{\partial x^{J}}\beta F\left(  x\right)  \right\vert
_{\widehat{\mathbf{x}}}+q_{\left(  a\right)  }^{I}\left(  \widehat{x}\right)
\widehat{\xi}_{\left(  a\right)  }\left(  t\right)  \label{sde}%
\end{equation}
where $\widehat{x}^{I}$, $I=1,...,N$ are the time-dependent stochastic
variables which are collectively denoted as $\widehat{x}$ (to simplify the
notation the time dependence is suppressed). The quantities $\widehat{\xi
}_{\left(  a\right)  }(t)$ are white, Gaussian noise delta-correlated in time
so that $\left\langle \widehat{\xi}_{\left(  a\right)  }\left(  t\right)
\right\rangle =0$ and $\left\langle \widehat{\xi}_{\left(  a\right)  }\left(
t_{1}\right)  \widehat{\xi}_{\left(  b\right)  }\left(  t_{2}\right)
\right\rangle =\delta_{\left(  ab\right)  }\delta\left(  t_{1}-t_{2}\right)
$. The number of noise terms may or may not be the same as the number of
variables and so is indexed differently with lower-case Roman characters as $1
\le a \le M$ where $M$. The Einstein summation convention is used according to
which all repeated indices are summed unless explicitly stated. To simplify
the notation, the definitions
\begin{align}
F_{I}^{\prime}\left(  y\right)   &  \equiv\left.  \frac{\partial}{\partial
x^{I}} \beta F\left(  x\right)  \right\vert _{y}\\
F_{IJ}^{\prime\prime}\left(  y\right)   &  \equiv\left.  \frac{\partial^{2}%
}{\partial x^{I}\partial x^{J}} \beta F\left(  x\right)  \right\vert
_{y}\nonumber
\end{align}
and $\det F^{\prime\prime}\left(  x\right)  $ for the determinant of the
second derivative matrix will be used. Thus, more compactly,
\begin{equation}
\frac{d}{dt}\widehat{x}^{I}=g^{IJ}\left(  \widehat{x}\right)  F_{I}^{\prime
}\left(  \widehat{x}\right)  +q_{\left(  a\right)  }^{I}\left(  \widehat
{x}\right)  \widehat{\xi}_{\left(  a\right)  }\left(  t\right)
\end{equation}
The first term on the right is the deterministic part of the dynamics and the
second models the stochastic forces. The matrix $g^{IJ}\left(  x\right)  $
goes by various names such as the kinetic coefficients or the diffusion matrix
and in general depends on the state of the system (i.e. on the instantaneous
values of the stochastic variables). The deterministic dynamics is driven by
the gradient of a potential or free energy (the notation suggests the latter
but either is possible). In the following, the existence of a
fluctuation-dissipation relation of the form $g^{IJ}\left(  x\right)
=q_{\left(  a\right)  }^{I}\left(  x\right)  q_{\left(  a\right)  }^{J}\left(
x\right)  $ will always be assumed. Because the prefactor of the Gaussian
noise is state-dependent, the stochastic differential equation is not fully
specified until a statistical interpretation is given\cite{Gardiner,Sokolov} as
discussed next.

The probability that the stochastic variables have a specific value at time
$t$, within some small volume $dx$ around the point $x$, is denoted $P\left(
x;t\right)  dx$ and it follows from the stochastic differential equation that
the probability density satisfies the Fokker-Plank equation\cite{Gardiner}%
\begin{equation}
\frac{\partial}{\partial t}P\left(  x;t\right)  =\mathcal{L}P\left(
x;t\right)
\end{equation}
with%

\begin{equation}
\mathcal{L}=\left\{
\begin{array}
[c]{c}%
-\frac{\partial}{\partial x^{I}}\left(  g^{IJ}\left(  x\right)  F_{J}^{\prime
}\left(  x\right)  +\frac{\partial}{\partial x^{J}}g^{IJ}\left(  x\right)
\right)  ,\;\;\text{(Ito)}\\
-\frac{\partial}{\partial x^{I}}\left(  g^{IJ}\left(  x\right)  F_{J}^{\prime
}\left(  x\right)  +q_{\left(  a\right)  }^{I}\left(  x\right)  \frac
{\partial}{\partial x^{J}}q_{\left(  a\right)  }^{J}\left(  x\right)  \right)
,\;\;\text{(Strat)}\\
-\frac{\partial}{\partial x^{I}}\left(  g^{IJ}\left(  x\right)  F_{J}^{\prime
}\left(  x\right)  +g^{IJ}\left(  x\right)  \frac{\partial}{\partial x^{J}%
}\right)  .\;\;\text{(anti-Ito)}%
\end{array}
\right.
\end{equation}
The particular form depends on the interpretation of the stochastic dynamics
given by Eq.(\ref{sde}) with the Ito interpretation favored in mathematical
developments and the Stratonovich, and  anti-Ito (also known in the literature as the H{\"a}nggi–Klimontovich) interpretations,
favored in physical applications (see, e.g. Sokolov\cite{Sokolov} for a recent, interesting discussion of the physical implications of each interpretation). Recall that they are equivalent in the sense
that, e.g., the Ito form can be written as
\begin{equation}
\mathcal{L}= -\frac{\partial}{\partial x^{I}}\left(  g^{IJ}\left(  x\right)
F_{J}^{\prime}\left(  x\right)  +\frac{\partial q_{\left(  a\right)  }%
^{I}\left(  x\right)  }{\partial x^{J}}q_{\left(  a\right)  }^{J}\left(
x\right)  +q_{\left(  a\right)  }^{I}\left(  x\right)  \frac{\partial
}{\partial x^{J}}q_{\left(  a\right)  }^{J}\left(  x\right)  \right)
\end{equation}
which is the Stratonovich form with an additional, non-conservative force in
the deterministic dynamics. Note that the FPE\ can be written in the more
compact and suggestive forms%

\begin{equation}
\mathcal{L}=\left\{
\begin{array}
[c]{c}%
-\frac{\partial}{\partial x^{I}}\left(  e^{-\beta F\left(  x\right)  }%
\frac{\partial}{\partial x^{J}}e^{\beta F\left(  x\right)  }g^{IJ}\left(
x\right)  \right)  ,\;\;\text{(Ito)}\\
-\frac{\partial}{\partial x^{I}}\left(  e^{-\beta F\left(  x\right)
}q_{\left(  a\right)  }^{I}\left(  x\right)  \frac{\partial}{\partial x^{J}%
}e^{\beta F\left(  x\right)  }q_{\left(  a\right)  }^{J}\left(  x\right)
\right)  ,\;\;\text{(Strat)}\\
-\frac{\partial}{\partial x^{I}}\left(  g^{IJ}\left(  x\right)  e^{-\beta
F\left(  x\right)  }\frac{\partial}{\partial x^{J}}e^{\beta F\left(  x\right)
}\right)  ,\;\;\text{(anti-Ito)}%
\end{array}
\right.
\end{equation}
from which one sees that the anti-Ito interpretation immediately implies a
stationary solution $P\left(  x;t\right)  \propto e^{-\beta F\left(  x\right)
}$. In heuristic models, this might be an important physical requirement (i.e.
that an equilibrium state exists corresponding to the free energy $F$) in
which case this serves to pick out the appropriate interpretation. In other
circumstances\cite{Moon_2014,Kupferman} in which the stochastic model is
derived from an underlying deterministic dynamics, the Stratonovich
interpretation is implied by the derivation making this a natural choice in
many heuristic models as well. The distinction between the different
interpretations obviously disappears when the noise amplitudes, and hence the
kinetic coefficients, are state-independent and furthermore, in some important
examples (see, e.g. \cite{Lutsko_JCP_2012_1,Lutsko_2025_1}) the anomalous
force evaluates to zero even though the amplitudes are state-dependent so that
the distinction between interpretations is irrelevant.

\subsection{Changes of variables and covariance}

If one makes a general change of coordinates, i.e. $y^{I}\left(  x\right)
\leftrightarrow x^{I}\left(  y\right)  $, and denotes the probability that the
$y$-variables have a given value at time $t$ as $\widetilde{P}\left(
y;t\right)  $ with the usual relation $\widetilde{P}\left(  y;t\right)
=\left(  \det\frac{\partial x\left(  y\right)  }{\partial y}\right)  P\left(
x\left(  y\right)  ;t\right)  $, and defining $\widetilde{F}\left(  y\right)
=F\left(  x\left(  y\right)  \right)  $ and $\widetilde{q}_{\left(  a\right)
}^{K}\left(  y\right)  =\frac{\partial y^{K}}{\partial x^{I}}q_{\left(
a\right)  }^{I}\left(  x\left(  y\right)  \right)  $ and $\widetilde{g}%
^{kl}\left(  y\right)  =\frac{\partial y^{K}}{\partial x^{I}}g^{IJ}\left(
x\left(  y\right)  \right)  \frac{\partial y^{L}}{\partial x^{J}},$the
Fokker-Planck equation for $P\left(  x;t\right)  $ becomes
\begin{equation}
\frac{\partial}{\partial t}\widetilde{P}\left(  y;t\right)  =-\frac{\partial
}{\partial y^{K}}\left(
\begin{array}
[c]{c}%
\widetilde{g}^{KL}\left(  y\right)  \frac{\partial\beta\widetilde{F}\left(
y\right)  }{\partial y^{L}}-\left(  \frac{\partial x^{I}}{\partial y^{R}}%
\frac{\partial x^{S}}{\partial y^{L}}\frac{\partial^{2}y^{K}}{\partial
x^{I}\partial x^{S}}\right)  \widetilde{g}^{RL}\left(  y\right)
+\frac{\partial}{\partial y^{L}}\widetilde{g}^{KL}\left(  y\right) \\
\widetilde{g}^{KL}\left(  y\right)  \frac{\partial\beta\widetilde{F}\left(
y\right)  }{\partial y^{L}}+\widetilde{q}_{\left(  a\right)  }^{K}\left(
y\right)  \frac{\partial}{\partial y^{L}}\widetilde{q}_{\left(  a\right)
}^{L}\left(  y\right) \\
\widetilde{g}^{KL}\left(  y\right)  \frac{\partial\left(  \beta\widetilde
{F}\left(  y\right)  +\ln\det\left(  \frac{\partial y}{\partial x}\right)
\right)  }{\partial y^{L}}+\widetilde{g}^{KL}\frac{\partial}{\partial y^{L}}%
\end{array}
\right)  \widetilde{P}\left(  y;t\right)  .
\end{equation}
The Stratonovich equation is manifestly covariant while in the anti-Ito
interpretation, the free energy is shifted by the log of the Jacobian
(corresponding to a shift of the equilibrium distribution from $e^{-\beta
F(x)}$ to $\det\left(  \frac{\partial x}{\partial y}\right)  e^{-\beta
\widetilde{F}(y)}$ as expected. In the Ito interpretation, the additional
contribution to the deterministic force, termed the anomalous force, is non-conservative.

\subsection{Barrier crossing and MFPT}

In general, multidimensional free energy surfaces could have all sorts of
complex or pathological topologies and a general treatment covering all
possibilities is likely impossible.\ For our purposes, it is assumed that the
free energy surface contains two or more local minima and can be divided into
two or more regions which which are each basins of attraction, under the
deterministic dynamics, for one of the minima. The basins of attraction are
separated from one another by a surface referred to as the separatrix. The
barrier crossing problem is defined by an two minima on the free energy
surface, $\mathbf{x}_{A}$ and $\mathbf{x}_{B}$, their associated basins of
attraction, denoted simply as $A$ and $B$, and the separatrix separating them,
$S_{AB}$. (Of course, two basins may not connect with on another but this is
then an example of the topologies beyond the present discussion). Any local
minimum on the separatrix is a saddle point and the lowest such saddle point
is the critical point. Here, attention is restricted to problems with a single
saddle point although generalization to multiple saddle points should not be difficult.

Given these restrictions, which are made in the classic analyses of the
problem, the barrier crossing problem asks for the mean first passage time for
a point that starts in the basin $A$, $x_{0}\in A$, to reach the separatrix.
This is denoted $T\left(  x_{0}\right)  $ and can be shown to be the solution
\begin{align}
\mathcal{L}^{\dag}T\left(  x\right)   &  =-1\\
T\left(  x\in S_{AB}\right)   &  =0\nonumber
\end{align}
where the first line involves the adjoint of the Fokker-Planck operator,%
\begin{equation}
\mathcal{L}^{\dag}=\left\{
\begin{array}
[c]{c}%
e^{\beta F\left(  x\right)  }g^{IJ}\left(  x\right)  \frac{\partial}{\partial
x^{J}}e^{-\beta F\left(  x\right)  }\frac{\partial}{\partial x^{I}%
},\;\;\text{(Ito)}\\
e^{\beta F\left(  x\right)  }q_{\left(  a\right)  }^{J}\left(  x\right)
\frac{\partial}{\partial x^{J}}q_{\left(  a\right)  }^{I}\left(  x\right)
e^{-\beta F\left(  x\right)  }\frac{\partial}{\partial x^{I}}%
,\;\;\text{(Strat)}\\
e^{\beta F\left(  x\right)  }\frac{\partial}{\partial x^{J}}g^{IJ}\left(
x\right)  e^{-\beta F\left(  x\right)  }\frac{\partial}{\partial x^{I}%
},\;\;\text{(anti-Ito)}%
\end{array}
\right. .
\end{equation}
In general, this is a second order equation and so requires a second boundary condition: for example, a common choice is to assume that the MFPT is independent of position far from the separatrix.

\subsection{The MLP\ and gradient descent pathways and distance measure}

In the Ito formulation, it is possible to give an explicit, exact expression
for the probability of any path between two given
points\cite{Onsager,Graham,Graham2}. This can in turn be maximized to
determine the most likely path (MLP) between the two points.\ If these are the
minima $\mathbf{x}_{A}$ and $\mathbf{x}_{B}$ this is the most likely
transition pathway.\ The general expressions are complicated but in the weak
noise approximation they become quite simple: the MLP\ passes through the
critical point and is determined by following the deterministic dynamics from
the critical point to the two minima.\ For a conservative force, this can be
interpreted as gradient descent starting at the critical point and falling
towards the two minima in a curved space with metric $g_{IJ}$ (defined as the
inverse of the matrix $g^{IJ}$). Explicitly, this is calculated by solving
\[
\frac{dx^{I}}{du}=g^{IJ}\left(  x\right)  \frac{\partial\beta F\left(
x\right)  }{\partial x^{J}},\;x\left(  0\right)  =x^{\ast}\pm\epsilon
\]
where $x^{\ast}$ is the critical point, $\epsilon$ is a small (infinitesimal)
perturbation in the unstable direction and $u$ is a parameter (not a physical
time). In this context, it is natural to measure distances in $x$-space using
this metric, so that the infinitesimal distance between points becomes%
\begin{equation}
ds=\sqrt{dx^{I}g_{IJ}\left(  x\right)  dx^{J}} \label{ds}%
\end{equation}
and the length of a parameterized path, given by $x^{I}\left(  u\right)  $ for
$u_{1}<u<u_{2}$, is
\begin{equation}
s=\int_{u_{1}}^{u_{2}}\sqrt{\frac{dx^{I}\left(  u\right)  }{du}g_{IJ}\left(
x\left(  u\right)  \right)  \frac{dx^{J}\left(  u\right)  }{du}}du.
\end{equation}
The distance between two points $x_{1}$ and $x_{2}$, is the minimum of this
path difference over all paths connecting the points.

Note that the parameter $u$ can be related to $s$ via
\begin{equation}
\frac{ds}{du}=\sqrt{\frac{dx^{I}}{du}g_{IJ}\left(  x\right)  \frac{dx^{J}}%
{du}}=\sqrt{F_{I}^{\prime}\left(  x\right)  g^{IJ}\left(  x\right)
F_{J}^{\prime}\left(  x\right)  }%
\end{equation}
so that it is then possible to eliminate the parameterization entirely and to
express the equation for the MLP in terms of this distance as%
\begin{equation}
\frac{dx^{I}}{ds}=\frac{du}{ds}\frac{\partial x^{I}}{\partial u}=\frac
{g^{IJ}\left(  x\right)  F_{J}^{\prime}\left(  x\right)  }{\sqrt{F_{K}%
^{\prime}\left(  x\right)  g^{KL}\left(  x\right)  F_{L}^{\prime}\left(
x\right)  }}%
\end{equation}
and observe that in one dimension this becomes%
\begin{equation}
\frac{dx}{ds}=\pm\frac{1}{\sqrt{\det g\left(  x\right)  }}=\pm q\left(
x\right)  .
\end{equation}

\section{Expressions for the MFPT}

\label{secMFPT}

\subsection{One dimension}

In one dimension, the free energy (or potential) is a function of a single
variable and the topology consists of two minima separated by a maximum. The
separatrix is therefore just the single point, denoted $x_{S}$, corresponding
to the critical point and, in this special case, it is in fact a local maximum
rather than a saddle point. Henceforth assume, without loss of generality,
that $x_{A}<x_{S}<x_{B}$. The MFPT equation in all interpretations can be
written as
\begin{align}
e^{\beta F\left(  x\right)  }q^{\alpha}\left(  x\right)  \frac{\partial
}{\partial x}q^{2-\alpha}\left(  x\right)  e^{-\beta F\left(  x\right)  }%
\frac{\partial}{\partial x}T\left(  x\right)   &  =-1\\
T\left(  x_{S}\right)   &  =0\nonumber\\
\lim_{x\rightarrow-\infty}\frac{d}{dx}T\left(  x\right)    &  =0\nonumber
\end{align}
with the values $\alpha=2,1,0$ being the Ito, Stratonovich and anti-Ito
interpretations, respectively. Note that the second boundary condition (the third line) is the heuristic requirement that the MFPT is independent of position far from the separatrix (critical point). An exact integral of the MFPT equation is then
\begin{equation}
\frac{\partial}{\partial x}T\left(  x\right)  =-q^{\alpha-2}\left(  x\right)
e^{\beta F\left(  x\right)  }\int_{-\infty}^{x}q^{-\alpha}\left(  z\right)
e^{-\beta F\left(  z\right)  }dz
\end{equation}
and a second integration gives%
\begin{equation}
T^{\left(  \text{Exact}\right)  }\left(  x_{A}\right)  =\int_{x_{A}}^{x_{S}%
}\;\frac{\int_{-\infty}^{y}q^{-\alpha}\left(  z\right)  e^{-\beta F\left(
z\right)  }dz}{q^{2-\alpha}\left(  y\right)  e^{-\beta F\left(  y\right)  }%
}dy. \label{Exact}%
\end{equation}
In the limit that $F\left(  y\right)  $ is sufficiently sharply peaked at
$y=x_{S}$, this can be approximated as%
\begin{equation}
T^{\left(  1\right)  }\left(  x_{A}\right)  \simeq\left(  \int_{x_{A}}^{x_{S}%
}\;\frac{1}{q^{2-\alpha}\left(  y\right)  e^{-\beta F\left(  y\right)  }%
}dy\right)  \left(  \int_{-\infty}^{x_{S}}q^{-\alpha}\left(  z\right)
e^{-\beta F\left(  z\right)  }dz\right)
\end{equation}
and a stationary phase approximation gives the classic Kramer formula%
\begin{equation}
\label{Kramer1D}T^{\left(  \text{Kramer}\right)  }\left(  x_{A}\right)
\simeq\left(  \pi\frac{q^{-\alpha}\left(  x_{A}\right)  }{q^{2-\alpha}\left(
x_{S}\right)  }\sqrt{\frac{1}{\left\vert \beta F^{\prime\prime}\left(
x_{S}\right)  \right\vert \beta F^{\prime\prime}\left(  x_{A}\right)  }%
}\right)  e^{\beta F\left(  x_{S}\right)  -\beta F\left(  x_{A}\right)  }%
\end{equation}

Finally, consider a general change of variables, $\widetilde{x}\left(
x\right)  \leftrightarrow x\left(  \widetilde{x}\right)  $, for which
\begin{equation}
T^{\left(  \text{Exact}\right)  }\left(  x_{A}\right)  =T^{\left(
\text{Exact}\right)  }\left(  \widetilde{x}\left(  x_{A}\right)  \right)
=\int_{\widetilde{x}_{A}}^{\widetilde{x}_{S}}\;\frac{\int_{-\infty
}^{\widetilde{y}}q^{-\alpha}\left(  z\left(  \widetilde{z}\right)  \right)
e^{-\beta\widetilde{F}\left(  \widetilde{z}\right)  }\frac{dz}{d\widetilde{z}%
}d\widetilde{z}}{q^{2-\alpha}\left(  \widetilde{y}\right)  e^{-\beta
\widetilde{F}\left(  \widetilde{y}\right)  }}\frac{dy}{d\widetilde{y}%
}d\widetilde{y}.
\end{equation}
It is interesting to note that if a change of variable is made to the natural
distance, $ds=\sqrt{dxg\left(  x\right)  dx}=\sqrt{g\left(  x\right)
}dx=q\left(  x\right)  dx$ then the MFPT\ becomes%
\begin{equation}
T^{\left(  \text{Exact}\right)  }\left(  x_{A}\right)  =\int_{s_{A}}%
^{0}\;\frac{\int_{-\infty}^{s}q^{1-\alpha}\left(  z\left(  s^{\prime}\right)
\right)  e^{-\beta F\left(  z\left(  s^{\prime}\right)  \right)  }ds^{\prime}%
}{q^{1-\alpha}\left(  x\left(  s\right)  \right)  e^{-\beta F\left(  x\left(
s\right)  \right)  }}ds
\end{equation}
and in particular, in the Stratonovich interpretation $\alpha=1$,
\begin{equation}
T^{\left(  \text{Exact}\right)  }\left(  x_{A}\right)  =\int_{s_{A}}%
^{0}\;e^{\beta F\left(  x\left(  s\right)  \right)  }\left(  \int_{-\infty
}^{s}e^{-\beta F\left(  x\left(  s^{\prime}\right)  \right)  }ds\right)
ds\;\text{Stratonovich,}%
\end{equation}
so that the state-dependence of the noise (i.e. the presence of the term
$q(x)$) is eliminated.

\subsection{Multidimensional}

One can try to follow the same steps for the multidimensional problem. To do
so, we introduce a series of surfaces $\Sigma\left(  s\right)  $ with the
following properties:

\begin{itemize}
\item The surface at $s=0$ (i.e. containing the critical point) is the
separatrix, $\Sigma\left(  0\right)  =S_{AB}$

\item The surface at $s_{A}$ contains the minimum, $x_{A}\in\Sigma\left(
s_{A}\right)  $.

\item The surfaces divide all of parameter space in two such that the minimum
$x_{A}$ is in one sub-volume, $V_{A}\left(  s\right)  $, and the separatrix
$S_{AB}$ and minimum $x_{B}$ is in the other sub-volume.

\item For all $0<s<s_{A}$, the MLP intersects $\Sigma\left(  s\right)  $ at
the single point $x\left(  s\right)  $.

\item The normal to the surface $\Sigma\left(  s\right)  $ at the point
$x\left(  s\right)  $ is \textbf{parallel to the path }$x(s)$\textbf{ and so
proportional to the gradient of the free energy, }$n_{I}\propto\beta
F_{I}^{\prime}\left(  x_{s}\right)  $\textbf{ except at the critical point
where it is proportional to the unstable left eigenvector.}
\end{itemize}

With various additional requirements (e.g. the surfaces never intersect one
another) the family of surfaces may form a foliation of the $x$-space but this
is not strictly required. Examples such families are shown in
Fig.\ \ref{surfaces}.

The details of how one proceeds depend on the exact form of the Fokker-Planck
operator, which is to say on the interpretation of the stochastic processes.
Since all interpretations are the same for the case of a constant
(state-independent) diffusion matrix, the various results should be degenerate
in this limit. Here, we begin with the anti-Ito interpretation which turns out
to be the most straightforward and which includes and from which the
state-independent limit follows immediately. This is followed by discussions
of the other two, more troublesome interpretations.

\subsubsection{Constant diffusion matrix and anti-Ito approximations}

Suppose that the surfaces $\Sigma\left(  s\right)  $ are parameterized in
terms of some set of variables $u_{a}$ for $1\leq a\leq N-1$ as $x^{\left(
s\right)  }\left(  u\right)  $ and that the point of intersection between
$\Sigma\left(  s\right)  $ and the MLP\ is the origin so that $x^{\left(
s\right)  }\left(  u=0\right)  =x\left(  s\right)  $. Integrating the MFPT
equation over $V\left(  s\right)  $ and using the divergence theorem gives%
\begin{align}
\int_{V_{A}\left(  s\right)  }e^{-\beta F\left(  x\right)  }dx  &
=\int_{V_{A}\left(  s\right)  }\frac{\partial}{\partial x^{J}}\left(
g^{IJ}\left(  x\right)  e^{-\beta F\left(  x\right)  }\frac{\partial}{\partial
x^{I}}T\left(  x\right)  \right)  \ d\mathbf{x}\\
&  =\int_{\Sigma\left(  s\right)  }g^{IJ}\left(  x^{\left(  s\right)  }\left(
u\right)  \right)  e^{-\beta F\left(  x^{\left(  s\right)  }\left(  u\right)
\right)  }\left(  \frac{\partial}{\partial x^{I}}T\left(  x\right)  \right)
_{x^{\left(  s\right)  }\left(  u\right)  }d\mathbf{\Sigma},\nonumber
\end{align}
where bold-face is used to indicate multi-dimensional measures (so that it is
easy to distinguish truly one-dimensional integrals and multidimensional
integrals) and it is assumed that any contribution from surfaces "at infinity"
are negligible. The surface element can be written explicitly in as
\begin{equation}
d\mathbf{\Sigma}_{J}=n_{J}\left(  u\right)  d\mathbf{u}%
\end{equation}
where $n_{J}\left(  u\right)  $ is normal to the surface, $n_{J}\left(
u\right)  \frac{\partial x^{J}\left(  u\right)  }{\partial u^{a}}=0$ and the
general expression for the normal is given in Eq.(\ref{normal}) with
particular instances being $\mathbf{n}=\pm\frac{\partial x^{2}}{\partial
u}\widehat{\mathbf{e}}_{1}\mp\frac{\partial x^{1}}{\partial u}\widehat
{\mathbf{e}}_{2}$ for a one-dimensional surface in two dimensions and the
familiar $\mathbf{n}=\pm\frac{\partial\mathbf{x}}{\partial u^{1}}\times
\frac{\partial\mathbf{x}}{\partial u^{2}}$ for a two-dimensional surface
embedded in 3 dimensions. The main approximation made now, having so chosen
the surfaces, is that the free energy restricted to the surface $\Sigma\left(
s\right)  $, $F\left(  x^{\left(  s\right)  }\left(  u\right)  \right)  $, is
sharply peaked at $u=0$ so that $x^{\left(  s\right)  }\left(  0\right)
=x\left(  s\right)  $ and
\begin{equation}
\int_{V_{A}\left(  s\right)  }e^{-\beta F\left(  x\right)  }d\mathbf{x}%
\simeq\left(  \frac{n_{I}^{\left(  s\right)  }\left(  0\right)  }{\sqrt
{\det\left(  \overline{g}^{\left(  s\right)  }\left(  0\right)  \right)  }%
}g^{IJ}\left(  x\left(  s\right)  \right)  \left(  \frac{\partial}{\partial
x^{J}}T\left(  x\right)  \right)  _{x\left(  s\right)  }\right)  \int
_{\Sigma\left(  s\right)  }e^{-\beta F\left(  x^{\left(  s\right)  }\left(
u\right)  \right)  }\sqrt{\det\left(  \overline{g}^{\left(  s\right)  }\left(
u^{\left(  s\right)  }\right)  \right)  }d\mathbf{u} \label{T0}%
\end{equation}
where the contribution at $\Sigma\left(  x_{A}\right)  $ has been dropped
under the assumption that $\frac{\partial}{\partial x^{I}}T\left(  x\right)  $
on this surface is negligible. We have also multiplied and divided by the
induced metric $\overline{g}_{ab}^{\left(  s\right)  }\left(  u\right)
=\frac{dx^{\left(  s\right)  I}\left(  u\right)  }{du^{a}}g_{IJ}\left(
x^{\left(  s\right)  }\left(  u\right)  \right)  \frac{dx^{\left(  s\right)
J}\left(  u\right)  }{du^{b}}$ in order that the surface integral is invariant
with respect to the choice of parmaterization. Further exact analysis yields,
given in Appendix \ref{Analysis} yields what will be called the "quasi-one
dimensional"\ approximation,
\begin{equation}
\label{Q1D}T^{\left(  Q1D\right)  }\left(  x_{A}\right)  \simeq\int_{0}%
^{s_{A}}\frac{\sqrt{\det\left(  g\left(  x\left(  s\right)  \right)  \right)
}\int_{V_{A}\left(  s\right)  }e^{-\beta F\left(  x\right)  }d\mathbf{x}}%
{\int_{\Sigma\left(  s\right)  }e^{-\beta F\left(  x^{\left(  s\right)
}\left(  u\right)  \right)  }\sqrt{\det\left(  \overline{g}^{\left(  s\right)
}\left(  u\right)  \right)  }d\mathbf{u}}ds
\end{equation}
and in fact the one-dimensional limit of this expression is
\begin{equation}
T\left(  x_{A}\right)  \simeq\int_{0}^{s_{A}}\frac{\sqrt{g\left(  x\left(
s\right)  \right)  }\int_{x\left(  s\right)  }^{x\left(  s_{A}\right)
}e^{-\beta F\left(  x\right)  }dx}{e^{-\beta F\left(  x\left(  s\right)
\right)  }}ds=\int_{x_{A}}^{x^{\ast}}\frac{g\left(  y\right)  \int_{y}^{x_{A}%
}e^{-\beta F\left(  x\right)  }dx}{e^{-\beta F\left(  y\right)  }}dy
\end{equation}
which reproduces the exact result of Eq.(\ref{Exact})\ in the anti-Ito
($\alpha=0$)\ case.

The integral in the denominator will be dominated by the smallest value of the
free energy on the surface $\Sigma\left(  s\right)  $ which, by hypothesis, is
the value on the MLP. The integral in the numerator will be dominated by the
smallest value of free energy in the volume $V\left(  s\right)  $, which by
hypothesis is $F\left(  x_{A}\right)  $. Taking these into account, one can
write this in a more revealing form as
\begin{equation}
T^{\left(  Q1D\right)  }\left(  x_{A}\right)  \simeq\int_{0}^{s_{A}}e^{\beta
F\left(  x\left(  s\right)  \right)  -\beta F\left(  x_{A}\right)  }%
\frac{\sqrt{\det\left(  g\left(  x\left(  s\right)  \right)  \right)  }%
\int_{V_{A}\left(  s\right)  }e^{-\left(  \beta F\left(  x\right)  -F\left(
x_{A}\right)  \right)  }d\mathbf{x}}{\int_{\Sigma\left(  s\right)
}e^{-\left(  \beta F\left(  x^{\left(  s\right)  }\left(  u\right)  \right)
-F\left(  x\left(  s\right)  \right)  \right)  }\sqrt{\det\left(  \overline
{g}^{\left(  s\right)  }\left(  u\right)  \right)  }d\mathbf{u}}ds.
\end{equation}
We now consider successively more drastic approximations to this expression.
First, one observes that, due to the exponential prefactor, the integral over
$s$ will be dominated by the values of $s$ for which $F\left(  x\left(
s\right)  \right)  $ is largest, namely those near the critical point. In this
region, it is to be expected that the integral over the volume $V_{A}\left(
s\right)  $ will be insensitive to the value of $s$ since most of its
contribution will come from the neighborhood of the attractor $x_{A}$ since it
is the region in which the integrand is the largest. This suggests the first
approximation
\begin{equation}
\label{T1}T^{\left(  1\right)  }\left(  x_{A}\right)  \simeq\left(  \int
_{0}^{s_{A}}\frac{e^{\beta F\left(  x\left(  s\right)  \right)  - \beta
F\left(  x_{A}\right)  }\sqrt{\det\left(  g\left(  x\left(  s\right)  \right)
\right)  }}{\int_{\Sigma\left(  s\right)  }e^{-\left(  \beta F\left(
\mathbf{x}^{\left(  s\right)  }\left(  \mathbf{u}\right)  \right)  -F\left(
\mathbf{x}\left(  s\right)  \right)  \right)  }\sqrt{\det\left(  \overline
{g}^{\left(  s\right)  }\left(  \mathbf{u}\right)  \right)  }d\mathbf{u}%
}ds\right)  \left(  \int_{V_{A}\left(  0\right)  }e^{-\left(  \beta F\left(
x\right)  - \beta F\left(  x_{A}\right)  \right)  }d\mathbf{x}\right)  .
\end{equation}

Next, the integral over the surface $\Sigma\left(  s\right)  $ is assumed to
be dominated by terms near the gradient descent path. Then it is reasonable to
make a stationary phase approximation for this integral which gives, see
Appendix \ref{StationaryPhase} for details,%
\begin{equation}
T^{\left(  2\right)  }\left(  x_{A}\right)  \simeq\left(  \left(  2\pi\right)
^{-\frac{n-1}{2}}\int_{0}^{s_{A}}\sqrt{\frac{\det M\left(  x\left(  s\right)
\right)  }{t^{J}\left(  x\left(  s\right)  \right)  F_{JK}\left(  x\left(
s\right)  \right)  t^{K}\left(  x\left(  s\right)  \right)  }}e^{\beta
F\left(  x\left(  s\right)  \right)  -\beta F\left(  x_{A}\right)  }ds\right)
\left(  \int_{V_{A}\left(  0\right)  }e^{-\left(  \beta F\left(  x\right)
-\beta F\left(  x_{A}\right)  \right)  }d\mathbf{x}\right)  \label{T2}%
\end{equation}
with the vector $t$ pointing along the path,
\begin{equation}
t^{I}\equiv\frac{dx^{I}}{ds}=\frac{g^{IJ}F_{J}}{\sqrt{F_{K}g^{KL}F_{L}}}%
\end{equation}
and
\begin{equation}
M_{IJ}=F_{IJ}-t_{I}t^{L}F_{LJ}-F_{IL}t^{L}t_{J}+2t_{I}t_{J}\left(  t^{L}%
F_{LM}t^{M}\right)  .
\end{equation}
It is tempting to insist on a similar approximation for the second factor on
the right, i.e. to say that%
\begin{align}
T^{\left(  3\right)  }\left(  x_{A}\right)   &  \simeq\left(  \int_{0}^{s_{A}%
}\sqrt{\frac{\det M\left(  x\left(  s\right)  \right)  }{t^{J}\left(  x\left(
s\right)  \right)  F_{JK}\left(  x\left(  s\right)  \right)  t^{K}\left(
x\left(  s\right)  \right)  }}e^{\beta F\left(  x\left(  s\right)  \right)
-\beta F\left(  x_{A}\right)  }ds\right)  \label{T3}\\
&  \times\left(  \int_{0}^{s_{A}}\sqrt{\frac{t^{J}\left(  x\left(  s\right)
\right)  F_{JK}\left(  x\left(  s\right)  \right)  t^{K}\left(  x\left(
s\right)  \right)  }{\det M\left(  x\left(  s\right)  \right)  }}e^{-\left(
\beta F\left(  x\left(  s\right)  \right)  -\beta F\left(  x_{A}\right)
\right)  }ds\right)  \nonumber
\end{align}
but this is only correct if every point in $V\left(  s=0\right)  =V_{A}$ but
not in $V\left(  s=s_{A}\right)  $ lies on one and only one of the family of
surfaces $\Sigma\left(  s\right)  $ and, while this may not in general be true
for any "natural" choice of surfaces it could perhaps be possible to impose it
as a second constraint on the surfaces (supplementing the constraint that the
surfaces be perpendicular to the GDP). As long as this is possible, the
approximation is viable. Note that this expression is actually amenable to
approximate numerical evaluation, even for high-dimensional problems since
efficient methods exist both for calculating the extreme (largest and
smallest) eigenvalues, as implemented in, e.g., the SLEPc
library\cite{SLEPC1,SLEPC2,SLEPC3} and as used e.g. in
\cite{Lutsko_Schoonen_JCP}, and also for evaluating high dimensional
determinants, see e.g. Refs.\cite{Det1,Det2}. A final level of approximation
is to perform stationary phase approximations for each of the two remaining
integrals in either Eq.(\ref{T2}) or \ref{T3} resulting in (see Appendix
\ref{StationaryPhase})
\begin{equation}
T^{\left(  \text{Langer}\right)  }\left(  x_{A}\right)  \simeq\pi\frac
{1}{\left\vert \lambda^{\left(  -\right)  }\right\vert }\frac{\sqrt
{\det\left\vert F^{\prime\prime}\left(  x^{\ast}\right)  \right\vert }}%
{\sqrt{\det F^{\prime\prime}\left(  x_{A}\right)  }}e^{\beta F(x^{\ast})-\beta
F(x_{A})}%
\end{equation}
Note that since $\lambda^{\left(  -\right)  }$ is the unstable eigenvalue of
the \emph{dynamical matrix}, in one dimensional one has that $\lambda
^{(-)}=g(x^{\ast})F^{\prime\prime}\left(  x^{\ast}\right)  $ and so the
one-dimensional limit of the the expression for the MFPT is
\begin{equation}
T^{\left(  \text{Langer}\right)  }\left(  x_{A}\right)  \simeq\frac{\pi
}{\left\vert g(x^{\ast})\right\vert \sqrt{\det\left\vert F^{\prime\prime
}\left(  x^{\ast}\right)  \right\vert }}\frac{1}{\sqrt{\det F^{\prime\prime
}\left(  x_{A}\right)  }}e^{\beta F(x^{\ast})-\beta F(x_{A})}%
\end{equation}
in agreement with the one-dimensional result, Eq.(\ref{Kramer1D}), for the
case $\alpha=0$. This expression is surprisingly simple, the only difference
from the constant diffusion tensor result being the appearance of the dynamic
eigenvalue. This single, but critical, change arises from the fact that the
surfaces are perpendicular to the unstable eigenvector of the dynamical matrix
at the critical point - the physically relevant quantities - and not the
unstable eigenvector derived from the free energy alone.

\subsubsection{Stratonovich interpretation}

The starting point in this case is the MFPT equation%
\begin{equation}
-e^{-\beta F\left(  x\right)  }=q_{\alpha}^{J}\left(  x\right)  \frac
{\partial}{\partial x^{J}}\left(  q_{\alpha}^{I}\left(  x\right)  e^{-\beta
F\left(  x\right)  }\frac{\partial}{\partial x^{I}}T\left(  x\right)  \right)
\
\end{equation}
where the number of noise terms, i.e. the range of the index $\alpha$ , is left
arbitrary. The difference from the previous case is that the right hand side
is no longer a perfect derivative. In one dimension, with a single noise
source, the noise amplitude becomes a scalar, $q_{\alpha}^{I}\left(  x\right)
\rightarrow q\left(  x\right)  $, and one simply divides by $q\left(
x\right)  $. In multiple dimensions, or with multiple noise terms, this is no
longer an option. One possibility is that $\frac{\partial}{\partial x^{J}%
}q_{\alpha}^{J}\left(  x\right)  =0$, which trivially includes the case of
state-independent amplitudes, the amplitude can simply be moved into the
gradient and the anti-Ito result applies.

Lacking this option, another possibility may be of some use and shows again
how similar the exact one-dimensional and approximate multi-dimensional
results can be. For example, restrict attention to the case that the number of
noise terms is the same as the number of dimensions and consider a general,
invertible change of variable described by $y\left(  x\right)
\longleftrightarrow x\left(  y\right)  $ for which the MFPT equation becomes
\begin{equation}
-e^{-\beta F\left(  x\left(  y\right)  \right)  }=q_{\alpha}^{J}\left(
x\left(  y\right)  \right)  \frac{\partial y^{K}}{\partial x^{J}}%
\frac{\partial}{\partial y^{K}}\left(  e^{-\beta F\left(  x\left(  y\right)
\right)  }q_{\alpha}^{I}\left(  x\left(  y\right)  \right)  \frac{\partial
y^{L}}{\partial x^{I}}\frac{\partial}{\partial y^{L}}T\left(  x\left(
y\right)  \right)  \right)  \
\end{equation}
and let us assume that we can find such a transformation that ensures that
\begin{equation}
q_{\alpha}^{J}\left(  x\left(  y\right)  \right)  \frac{\partial y^{K}%
}{\partial x^{J}}=\overline{q}_{\alpha}^{K}%
\end{equation}
where the matrix on the right hand side is constant. Without loss of
generality, assume it is diagonal and after multiplying through by
$g_{JI}\left(  x\right)  q_{\alpha}^{I}\left(  x\right)  q_{\alpha}^{I}\left(
x\right)  $ one must solve%
\begin{equation}
\frac{\partial y^{K}}{\partial x^{J}}=g_{JI}\left(  x\right)  q^{IK}\left(
x\right)
\end{equation}
and this is only possible if and only if%
\begin{equation}
\frac{\partial}{\partial x^{L}}\left(  g_{JI}\left(  x\right)  q^{IK}\left(
x\right)  \right)  =\frac{\partial}{\partial x^{J}}\left(  g_{LI}\left(
x\right)  q^{IK}\left(  x\right)  \right)  .
\end{equation}
This then gives
\begin{equation}
-e^{-\beta\widetilde{F}\left(  y\right)  }=\frac{\partial}{\partial y^{K}%
}\left(  e^{-\beta\widetilde{F}\left(  y\right)  }\overline{g}^{KL}%
\frac{\partial}{\partial y^{L}}\widetilde{T}\left(  y\right)  \right)  \
\end{equation}
and from the anti-Ito results one immediately finds
\begin{align}
T\left(  x_{A}\right)   &  \simeq\int_{0}^{s_{A}}\frac{\sqrt{\det\left(
\overline{g}\right)  }\int_{V_{A}\left(  s\right)  }e^{-\beta\widetilde
{F}\left(  y\right)  }d\mathbf{y}}{\int_{\Sigma\left(  s\right)  }%
e^{-\beta\widetilde{F}\left(  y^{\left(  s\right)  }\left(  u\right)  \right)
}\sqrt{\det\left(  \overline{\overline{g}}^{\left(  s\right)  }\left(
u\right)  \right)  }d\mathbf{u}}ds\\
&  =\int_{0}^{s_{A}}\frac{\sqrt{\det\left(  \overline{g}\right)  }\int
_{V_{A}\left(  s\right)  }e^{-\beta F\left(  y\right)  }\det\left(
gq_{\alpha}\overline{q}_{\alpha}\right)  d\mathbf{x}}{\int_{\Sigma\left(
s\right)  }e^{-\beta F\left(  x^{\left(  s\right)  }\left(  u\right)  \right)
}\sqrt{\det\left(  \overline{g}^{\left(  s\right)  }\left(  u\right)  \right)
}d\mathbf{u}}ds\nonumber\\
&  =\int_{0}^{s_{A}}\frac{\int_{V_{A}\left(  s\right)  }e^{-\beta F\left(
x\right)  }\sqrt{\det g\left(  \mathbf{x}\right)  }d\mathbf{x}}{\int
_{\Sigma\left(  s\right)  }e^{-\beta F\left(  x^{\left(  s\right)  }\left(
u\right)  \right)  }\sqrt{\det\left(  \overline{g}^{\left(  s\right)  }\left(
u\right)  \right)  }d\mathbf{u}}ds\nonumber
\end{align}
and in the one-dimensional limit,%
\begin{equation}
T\left(  x_{A}\right)  \simeq\int_{0}^{s_{A}}\frac{\int_{x\left(  s\right)
}^{x\left(  s_{A}\right)  }e^{-\beta F\left(  x\right)  }\frac{1}{\det
q\left(  x\right)  }dx}{e^{-\beta F\left(  x\left(  s\right)  \right)  }%
}ds=\int_{0}^{y_{A}}\frac{\sqrt{g\left(  y\right)  }\int_{y}^{y_{A}}e^{-\beta
F\left(  x\right)  }\frac{1}{\det q\left(  x\right)  }dx}{e^{-\beta F\left(
x\left(  s\right)  \right)  }}dy
\end{equation}
which is the exact one-dimensional result for the Stratonovich interpretation.

If the family of surfaces $\Sigma\left(  s\right)  $ forms a foliation of the
parameter space, one can write
\begin{equation}
\int_{V_{A}\left(  s\right)  }e^{-\beta F\left(  x\right)  }\sqrt{\det
g\left(  \mathbf{x}\right)  }d\mathbf{x=}\int_{-\infty}^{s}\left(
\int_{\Sigma\left(  s\right)  }e^{-\beta F\left(  x\right)  }\sqrt{\det\left(
\overline{g}^{\left(  s\right)  }\left(  u\right)  \right)  }d\mathbf{u}%
\right)  ds
\end{equation}
where use has been made of the fact that the surfaces are perpendicular to
$\frac{\partial\mathbf{x}}{\partial s}$ and that $ds^{2}=dx^{I}g_{IJ}dx^{J}$.
Hence%
\begin{equation}
T\left(  x_{A}\right)  \simeq\int_{0}^{s_{A}}e^{\beta\mathcal{F}\left(
s\right)  }\left(  \int_{s}^{\infty}e^{-\beta\mathcal{F}\left(  s^{\prime
}\right)  }ds^{\prime}\right)  ds
\end{equation}
with%
\begin{equation}
e^{-\beta\mathcal{F}\left(  s\right)  }\equiv\frac{\int_{\Sigma\left(
s\right)  }e^{-\beta F\left(  x^{\left(  s\right)  }\left(  u\right)  \right)
}\sqrt{\det\left(  \overline{g}^{\left(  s\right)  }\left(  u\right)  \right)
}d\mathbf{u}}{\int_{0}^{\infty}\left(  \int_{\Sigma\left(  s^{\prime}\right)
}e^{-\beta F\left(  x^{\left(  s^{\prime}\right)  }\left(  u\right)  \right)
}\sqrt{\det\left(  \overline{g}^{\left(  s\right)  }\left(  u\right)  \right)
}d\mathbf{u}\right)  ds^{\prime}}%
\end{equation}

\bigskip

\bigskip

\section{Numerical Experiments}

\label{secNumerical}

\begin{table}%
\begin{tabular}
[c]{|c|cccc|}
& \multicolumn{4}{c|}{Gauss1}\\\hline
$i$ & $1$ & $2$ & $3$ & $4$\\\hline
$A_{i}$ & $-2$ & $1.5$ & $0.1$ & $-4$ \\
$x_{i}$ & $-1$ & $0$ & $0$ & $1$\\
$y_{i}$ & $0$ & $0$ & $0$ & $0$\\
$a$ & $-1$ & $-1$ & $0.25$ & $-1$\\
$b$ & - & - & - & -\\
$c$ & $-1$ & $0.5$ & $0.25$ & $-1$%
\end{tabular}
%}
\caption{Coefficients used to define the Gauss1 potential. Gauss2 is the same as Gauss1 except the $y_1 = -1$.}
\label{table1}
\end{table}

\begin{figure}
[ptb]\includegraphics[width=0.45\linewidth]{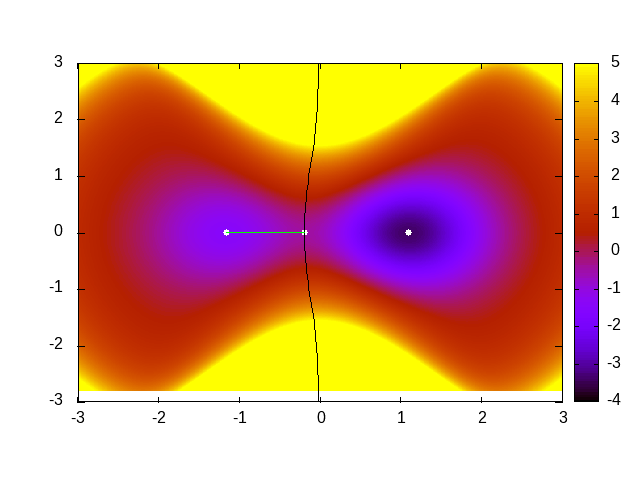}
\includegraphics[width=0.45\linewidth]{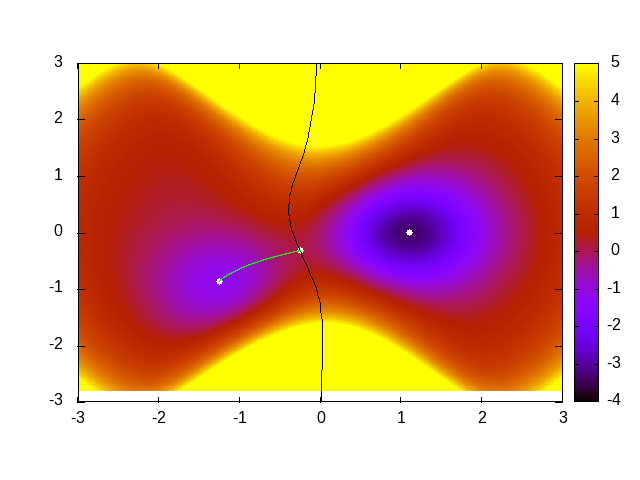}
\caption{Heat maps of the potential surfaces for two potentials, Gauss1 on the left and Gauss2 on the right. The stationary points are shown as white spots, the gradient descent pathway connecting the saddle point to the left minimum is a green line and the black line also passing through the saddle point is the separatrix (all calculated using the Diagonal metric).}
\label{surfaces}
\end{figure}

To test these expressions, we present a series of numerical experiments in two
dimensions. The potentials used are of the Muller-Brown form,
\begin{equation}
V(x,y)= \sum_{i=1}^{4}A_{i}\exp\left(  a_{i}\left(  x-x_{i}\right)  ^{2}%
+b_{i}\left(  x-x_{i}\right)  \left(  y-y_{i}\right)  +c_{i}\left(
y-y_{i}\right)  ^{2}\right)
\end{equation}
with parameters given in Table \ref{table1} and the Gauss2 potential is the
same as the Gauss 1 except that $y_{1} = -1$. Heat maps of potential surfaces
are shown in Fig. \ref{surfaces}. The metrics are specified in terms of the
noise amplitude, $q$, since these can be arbitrary while assuring that the
actual metrics are always symmetric and positive semi-definite, as required.
The general form used is
\begin{equation}
q\left(  x,y\right)  =\left(
\begin{array}
[c]{cc}%
q_{11}\times\left(  1+A_{11}\sin\left(  8\pi x\right)  \right)  & q_{12}\\
q_{21} & q_{22}\times\left(  1+A_{22}\sin\left(  8\pi y\right)  \right)
\end{array}
\right)
\end{equation}
with the following cases:

\begin{itemize}
\item Diagonal:\ $q_{11}=q_{22}=1$, all other constants are zero

\item NonDiagonal:\ $q_{11}=0.75$, $q_{12}=0.25$, $q_{22}=1.3$ and all other
constants zero

\item Variable:\ same as NonDiagonal with $A_{11}=A_{22}=0.5$
\end{itemize}

\begin{figure}
[ptb]\includegraphics[width=0.45\linewidth]{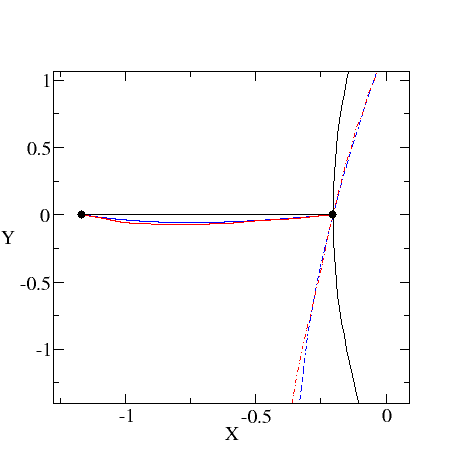}
\includegraphics[width=0.45\linewidth]{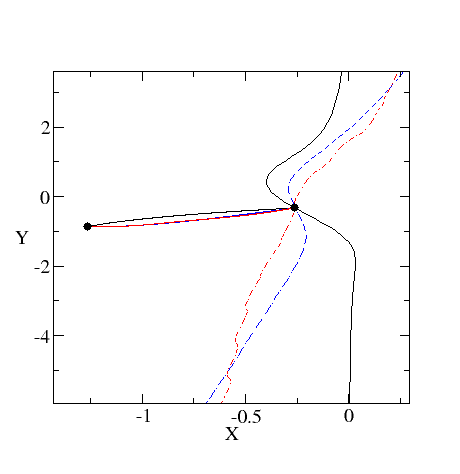}
\caption{GDP and separatrix for the Gauss1 (left) and Gauss2 (right) potentials with different metrics: the Diagonal metric (black, full lines), the NonDiagonal metric (blue, dashed lines) and the Variable metric (red, dot-dashed lines).
} \label{qmetric}
\end{figure}

The choice of metric does not, of course, affect the free energy surfaces but
does play a role in determining the GDP\ and the separatrix. The resulting
curves are shown in Fig. \ref{qmetric} where it is seen that the effect of the
various metrics on the GDP are small for both potentials, while the effect of
the non-diagonal metric on the separatrix is much larger and that of the
variability is only significant for the Gauss2 case.

\begin{figure}
[ptb]\includegraphics[width=0.45\linewidth]{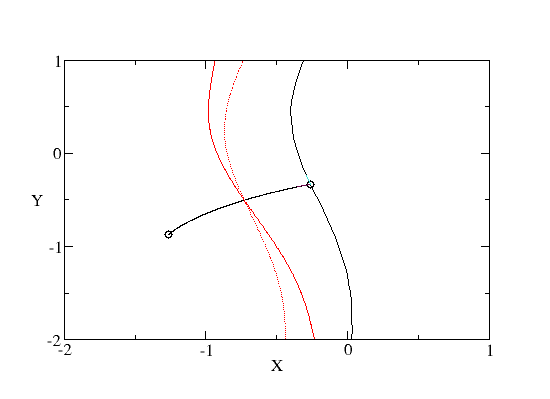}
\includegraphics[width=0.45\linewidth]{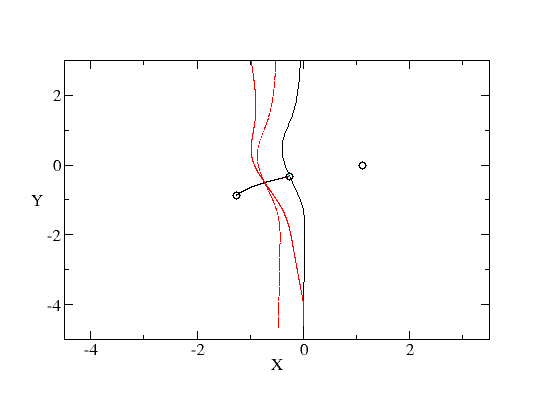}
\caption{Illustration of the ``dragging'' of the separatrix from the critical point to the left minimum (using the Gauss2 potential). The dotted line is a copy of the separatrix shifted so that the point of intersection of the gradient descent path is the same as at the saddle point while the full line is the same curve rotated so that its normal at the intersection is parallel to that of the pathway. The right panel shows how the curve extends until it intersects the original separatrix at which point the latter is used.}
\label{drag}
\end{figure}

To evaluate the various quasi-1D approximations, a set of surfaces must be
defined. The choice used here is to drag the separatrix along the gradient
descent path while rotating it so that it is always perpendicular to the GDP
at the point that the two intersect. Any time the resulting surface (line in
two dimensions) would cross the separatrix, the separatrix is used, as
illustrated in Fig. \ref{drag}.

The results of evaluating the quasi-1D approximations are shown in Fig.
\ref{results1} for the Gauss1 potential. Also shown are simulation results
obtained as discussed in Appendix \ref{Simulations}. In all cases, the
quantity displayed is the $Z \equiv T e^{-(\beta\Delta V)}$ where $T$ is the
mean first passage time and $\Delta V$ is the energy barrier. For the
topologically simplest case, it is seen that the quasi-1D approximation with
the separatrix-based surfaces agrees very well with the simulation results
over the entire range of temperatures. Not surprisingly, there is little
difference when using the planar surfaces. The two intermediate
approximations, $T^{(1)}$ and $T^{(2)}$ capture the general trend but
over-estimate the magnitude of $Z$. In contrast, the approximations $T^{(3)}$
and $T^{(4)}$, in which one or both of the multidimensional integrals is
evaluated using the stationary phase approximation, give rather poor results.

\begin{figure}
[ptb]\includegraphics[width=0.3\linewidth]{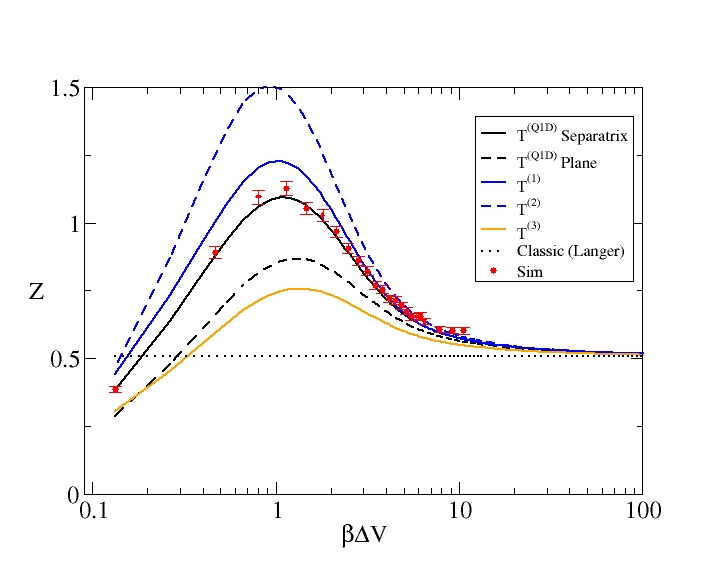}
\includegraphics[width=0.3\linewidth]{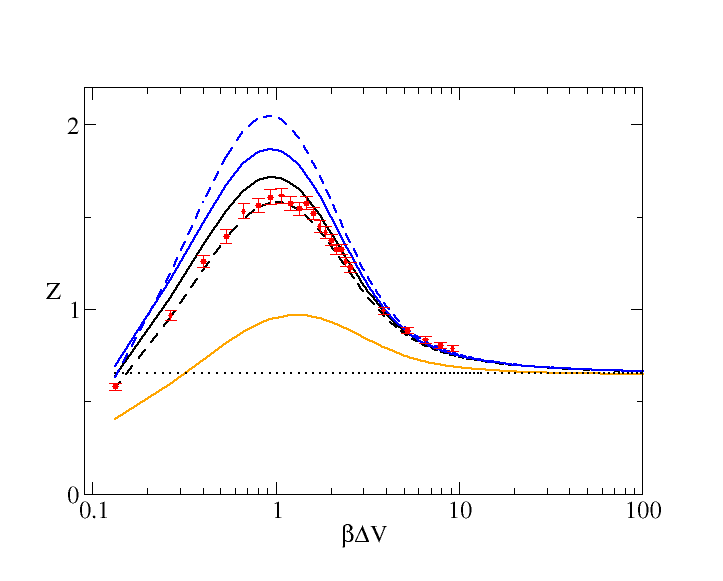}
\includegraphics[width=0.3\linewidth]{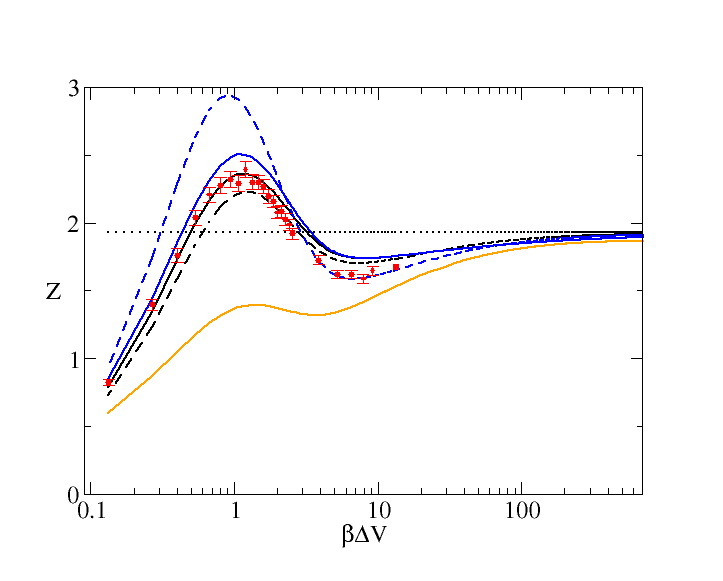}
\caption{MFPT expressed in terms of the coefficient of the exponential term for the Gauss2 with the Diagonal, NonDiagonal and Variable metrics (left, center and right panels, respectively). Note that the calculations with the variable metric are extended to $\beta \Delta V = 500$ in order to capture the convergence to the asymptotic (Langer) result. The Q1D approximation has been evaluated for two cases: one in which the foliation is a result of dragging the separatrix, as described in the text, and a second in which the separatrix has been replaced to a plane that passes through the critical point where it is perpendicular to the unstable eigenvector.}
\label{results2}
\end{figure}

Figure \ref{results2} show similar results using the topologically more
complicated Gauss2 potential. In this case, the separatrix-based quasi-1D
approximation is again quite good for the identity metric, becomes slightly
worse for the non-diagonal metric and degrades further for the variable
metric. Nevertheless, the comparisons in all three cases are reasonable. The
trends with the various approximations are similar to the Gauss1 results.

\begin{figure}
[ptb]\includegraphics[width=0.3\linewidth]{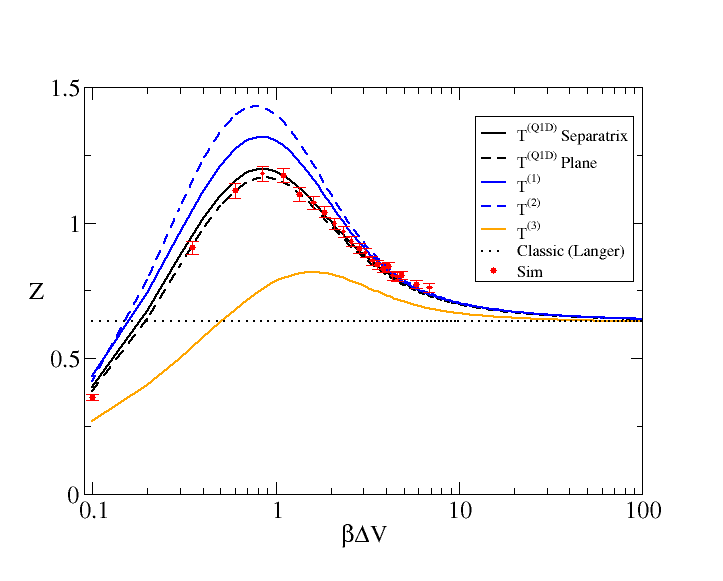}
\includegraphics[width=0.3\linewidth]{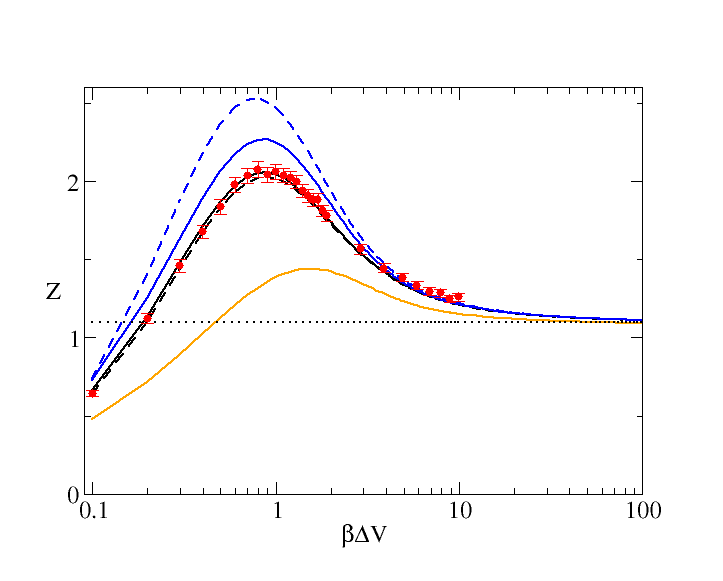}
\includegraphics[width=0.3\linewidth]{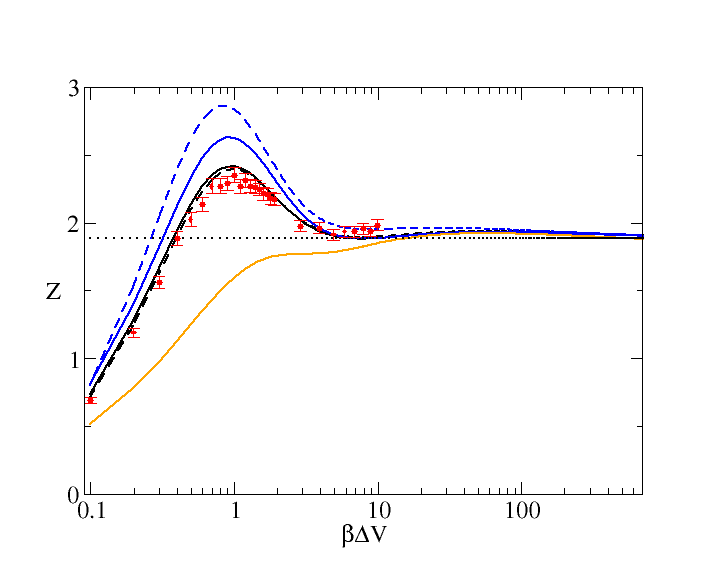}
\caption{MFPT expressed in terms of the coefficient of the exponential term for the Gauss1 with the Diagonal, NonDiagonal and Variable metrics (left, center and right panels, respectively). Note that the calculations with the variable metric are extended to $\beta \Delta V = 500$ in order to capture the convergence to the asymptotic (Langer) result. The Q1D approximation has been evaluated for two cases: one in which the foliation is a result of dragging the separatrix, as described in the text, and a second in which the separatrix has been replaced to a plane that passes through the critical point where it is perpendicular to the unstable eigenvector.}
\label{results1}
\end{figure}

\section{Conclusions}

In this work, the intuitively appealing idea that stochastically driven
multi-dimensional barrier-crossing processes can be approximated by
one-dimensional models has been studied. It was assumed that the topology of
the free energy surface is ``sufficiently'' simple that it makes sense to
think it in terms of the basins of attraction of two energy minima separated
by a critical point (the minimum barrier on the separatrix). A family of
surfaces along the steepest descent path connecting the critical point to the
starting minimum was defined by the requirement that each surface be
perpendicular to the steepest descent path with the initial surface being the
separatrix itself. The exact equation for the mean first passage time has
reduced to an effective one-dimensional equation by assuming that the rate of
change of the MFPT as one moves away from the separatrix towards the starting
minimum varies slowly in directions perpendicular to the steepest descent
path (i.e. is slowly varying on the surfaces defined above). This gives
Eq.(\ref{Q1D}) which is an effectively one-dimensional model for the MFPT and
which reduces to the exact expression in one-dimension. Various further
simplifying approximations were discussed culminating in the classic
expression of Langer\cite{Langer1,Langer2}.

The various approximations were evaluated using two-dimensional problems with
diagonal, non-diagonal and non-constant,non-diagonal metrics and compared to
simulations of the stochastic processes. It was found that the most general,
so-called ``quasi-one dimensional'', approximation gives surprisingly good
results for the systems studied, capturing quantitatively most of the
variation of non-Arrhenius dependence of the MFPT on the temperature. The
various simpler, more tractable, approximations give increasingly worse
results as more approximations are introduced until the classic expression
predicts no non-Arrhenius behavior at all, thus missing all such structure.

The approximations studied here could be useful in practical problems for
which the assumptions of the classical approximation are too crude: e.g. for
nucleation when the energy barrier is not large compared to the temperature.
The results given here then provide at least a baseline for anticipating how
much such a calculation should be trusted (or not). It would be interesting in
the future to extend this baseline by performing similar studies in higher dimensions.

\begin{acknowledgments}
This work was supported by the European Space Agency (ESA) and the Belgian
Federal Science Policy Office (BELSPO) in the framework of the PRODEX
Programme, Contract No. ESA AO-2004-070.
\end{acknowledgments}

\bibliography{./mnt.bib}

\appendix

\section{Derivation of Eq.(\ref{Q1D})}

\label{Analysis}

\subsection{The main result}

We begin with Eq.(\ref{T1}) of the main text, repeated here for convenience,
\begin{equation}
\label{T11}\int_{V\left(  s\right)  }e^{-\beta F\left(  x\right)  }%
d\mathbf{x}\simeq\left(  \frac{n_{I}^{\left(  s\right)  }\left(  0\right)
}{\sqrt{\det\left(  \overline{g}^{\left(  s\right)  }\left(  0\right)
\right)  }}g^{IJ}\left(  x\left(  s\right)  \right)  \left(  \frac{\partial
}{\partial x^{J}}T\left(  x\right)  \right)  _{x\left(  s\right)  }\right)
\int_{\Sigma\left(  s\right)  }e^{-\beta F\left(  x^{\left(  s\right)
}\left(  u\right)  \right)  }\sqrt{\det\left(  \overline{g}^{\left(  s\right)
}\left(  u^{\left(  s\right)  }\right)  \right)  }d\mathbf{u}.
\end{equation}

By hypothesis, the normal of the surface $\Sigma\left(  s\right)  $ at the
point $x\left(  s\right)  =x\left(  u=0\right)  $ is proportional to the
energy gradient, $n_{I}\left(  0\right)  =\gamma\beta F_{I}^{\prime}\left(
x_{s}\right)  $ for some constant $\gamma$. Contracting this first with
$g^{IJ}n_{J}$ and then with $g^{IJ}\beta F_{J}^{\prime}\left(  x_{s}\right)  $
and dropping all of the arguments for clarity, gives
\begin{equation}
\gamma=\frac{n_{I}g^{IJ}n_{J}}{F_{K}^{\prime}g^{KM}n_{M}}=\frac{n_{I}%
g^{IJ}F_{J}^{\prime}}{F_{K}^{\prime}g^{KM}F_{M}^{\prime}}%
\end{equation}
which can be combined into
\begin{equation}
\gamma=\frac{\sqrt{n_{I}g^{IJ}n_{J}}}{\sqrt{F_{K}^{\prime}g^{KM}F_{M}^{\prime
}}}.
\end{equation}
Using this, one sees that
\begin{align}
\frac{n_{I}^{\left(  s\right)  }}{\sqrt{\det\left(  \overline{g}^{\left(
s\right)  }\right)  }}g^{IJ}\frac{\partial}{\partial x^{J}}T  &  =\frac
{1}{\sqrt{\det\left(  \overline{g}^{\left(  s\right)  }\right)  }}\left(
\frac{\sqrt{n_{I}g^{IJ}n_{J}}}{\sqrt{F_{K}^{\prime}g^{KM}F_{M}^{\prime}}}\beta
F_{I}^{\prime}\right)  g^{IJ}\frac{\partial}{\partial x^{J}}T\\
&  =\frac{\sqrt{n_{I}g^{IJ}n_{J}}}{\sqrt{\det\left(  \overline{g}^{\left(
s\right)  }\right)  }}\frac{d}{ds}T\nonumber
\end{align}
so that Eq.(\ref{T11}) can be rearranged as
\begin{align}
\frac{d}{ds}T\left(  x\left(  s\right)  \right)   &  \simeq\frac{\sqrt
{\det\left(  \overline{g}\left(  u=0\right)  \right)  }}{\sqrt{n_{I}\left(
0\right)  g^{IJ}\left(  x\left(  s\right)  \right)  n_{J}\left(  0\right)  }%
}\frac{\int_{V\left(  s\right)  }e^{-\beta F\left(  x\right)  }d\mathbf{x}%
}{\int_{\Sigma\left(  s\right)  }e^{-\beta F\left(  x^{\left(  s\right)
}\left(  u\right)  \right)  }\sqrt{\det\left(  \overline{g}^{\left(  s\right)
}\left(  u\right)  \right)  }d\mathbf{u}}\\
&  =\frac{\sqrt{\det\left(  g\left(  x\left(  s\right)  \right)  \right)
}\int_{V\left(  s\right)  }e^{-\beta F\left(  x\right)  }dx}{\int
_{\Sigma\left(  s\right)  }e^{-\beta F\left(  x^{\left(  s\right)  }\left(
u\right)  \right)  }\sqrt{\det\left(  \overline{g}^{\left(  s\right)  }\left(
u\right)  \right)  }d\mathbf{u}}\nonumber
\end{align}
where the final line follows from a straightforward, but technical, argument
given in the subsection below. Integrating and using the boundary condition on
the separatrix, $T\left(  x\in S_{AB}\right)  =0$, gives the final result
\begin{equation}
T\left(  x_{A}\right)  \simeq\int_{0}^{s_{A}}\frac{\sqrt{\det\left(  g\left(
x\left(  s\right)  \right)  \right)  }\int_{V\left(  s\right)  }e^{-\beta
F\left(  x\right)  }d\mathbf{x}}{\int_{\Sigma\left(  s\right)  }e^{-\beta
F\left(  x^{\left(  s\right)  }\left(  u\right)  \right)  }\sqrt{\det\left(
\overline{g}^{\left(  s\right)  }\left(  u\right)  \right)  }d\mathbf{u}}ds.
\end{equation}

\subsection{Proof of identity}

In order to set the notation we first recall some basic results from linear
algebra. For a general $d$-dimensional square matrix $m_{ij}$, the minor,
$M\left(  m\right)  _{ij}$ is the determinant of the matrix formed by removing
the $i$th row and $j$th column. The cofactor matrix, $C\left(  m\right)
_{ij}=\left(  -1\right)  ^{I+J}M\left(  m\right)  _{ij}$ , the determinant is
$\det\left(  m\right)  =\sum_{i=1}^{D}m_{ij}C\left(  m\right)  _{ij}$ and the
inverse of $m$ is $m^{-1}=\frac{1}{\det m}C\left(  m\right)  ^{T}=\frac
{1}{\det m}A\left(  m\right)  $, where $A\left(  m\right)  =C\left(  m\right)
^{T}$. Note that for any two matrices $m_{1}$ and $m_{2}$
\begin{align}
A\left(  m_{1}m_{2}\right)   &  =A\left(  m_{2}\right)  A\left(  m_{1}\right)
\\
&  \rightarrow C^{T}\left(  m_{1}m_{2}\right)  =C^{T}\left(  m_{2}\right)
C^{T}\left(  m_{1}\right) \nonumber\\
&  \rightarrow C\left(  m_{1}m_{2}\right)  =C\left(  m_{1}\right)  C\left(
m_{2}\right)  .\nonumber
\end{align}

For an $n-1$ dimensional surface embedded in an $n$ dimensional space, let the
coordinates of the latter be $x^{I}$ and the surface be parameterized with
$u^{\left(  a\right)  }$, so $x=x\left(  u\right)  $ on the surface, and the
tangent vectors at the point $u_{0}$ are $t_{a}^{I}=\frac{\partial
x^{I}\left(  u\right)  }{\partial u^{a}}$. Defining the matrix%
\begin{equation}
N=\left(
\begin{array}
[c]{cccc}%
1 & 1 & \ldots & 1\\
\frac{dx^{1}}{du^{1}} & \frac{dx^{2}}{du^{1}} & \ldots & \frac{dx^{n}}{du^{1}%
}\\
& \vdots &  & \\
\frac{dx^{1}}{du^{n-1}} & \frac{dx^{2}}{du^{n-1}} & \ldots & \frac{dx^{n}%
}{du^{n-1}}%
\end{array}
\right)  =\left(
\begin{array}
[c]{c}%
\overrightarrow{1}_{n}\\
\overrightarrow{t}_{1}\\
\vdots\\
\overrightarrow{t}_{n}%
\end{array}
\right)  \label{normal}%
\end{equation}
one sees that the vector $n_{I}=C\left(  N\right)  _{1I}$ is orthogonal to the
tangent vectors since%
\begin{equation}
n_{I}t_{a}^{I}=C\left(  N\right)  _{1I}t_{a}^{I}=\det\left(
\begin{array}
[c]{c}%
\overrightarrow{t}_{a}\\
\overrightarrow{t}_{1}\\
\vdots\\
\overrightarrow{t}_{n}%
\end{array}
\right)  =0
\end{equation}
since the matrix has repeated rows.

Our goal is to prove the claimed equivalence%
\begin{equation}
\frac{\det\left(  \overline{g}\left(  \mathbf{u}_{0}^{\left(  s\right)
}\right)  \right)  }{n_{I}\left(  \mathbf{u}_{0}^{\left(  s\right)  }\right)
g^{IJ}\left(  x\left(  \mathbf{u}_{0}^{\left(  s\right)  }\right)  \right)
n_{J}\left(  \mathbf{u}_{0}^{\left(  s\right)  }\right)  }=\det\left(
g\left(  x\left(  \mathbf{u}_{0}^{\left(  s\right)  }\right)  \right)
\right)  ,
\end{equation}
where $\overline{g}_{ab}=t_{a}^{I}g_{IJ}t_{b}^{J}$. Dropping all arguments
since everything is evaluated at a fixed point, consider
\begin{align}
n_{I}g^{IJ}n_{J}\times\det g  &  =C\left(  N\right)  _{1I}\left(  \frac
{1}{\det g}C\left(  g\right)  _{IJ}\right)  C\left(  N\right)  _{1J}\times\det
g\\
&  =\left(  -1\right)  ^{1+I+I+J+1+J}M\left(  N\right)  _{1I}M\left(
g\right)  _{IJ}M\left(  N\right)  _{1J}\nonumber\\
&  =M\left(  NgN^{T}\right)  _{11}.\nonumber
\end{align}
Noting that
\begin{align}
NgN^{T}  &  =N\left(
\begin{array}
[c]{cccc}%
g_{IK}1^{K} & g_{IK}t_{1}^{K} & ... & g_{IK}t_{n-1}^{K}%
\end{array}
\right) \\
&  =\left(
\begin{array}
[c]{cccc}%
1^{I}g_{IK}1^{K} & 1^{I}g_{IK}t_{1}^{K} & \ldots & 1^{I}g_{IK}t_{n-1}^{K}\\
t_{1}^{I}g_{IK}1^{K} & t_{1}^{I}g_{IK}t_{1}^{K} & \ldots & t_{1}^{I}%
g_{IK}t_{n-1}^{K}\\
& \vdots &  & \\
t_{n-1}^{I}g_{IK}1^{K} & t_{n-1}^{I}g_{IK}t_{1}^{K} & \ldots & t_{n-1}%
^{I}g_{IK}t_{n-1}^{K}%
\end{array}
\right) \nonumber
\end{align}
one sees
\begin{equation}
M\left(  NgN^{T}\right)  _{11}=\det\left(  t_{a}^{I}g_{IK}t_{b}^{K}\right)
\equiv\det\overline{g}%
\end{equation}
thus proving the equivalence. It should be noted that this only proves the
identity when the normal is the vector given by Eq.(\ref{normal}) and not its
normalized form.

\section{Stationary phase approximation\label{StationaryPhase}}

\qquad We need to approximate Eq.(\ref{T1}), which can for convenience is
reproduced here,
\begin{equation}
\label{T10}T^{\left(  1\right)  }\left(  x_{A}\right)  \simeq\left(  \int
_{0}^{s_{A}}\frac{e^{\beta F\left(  x\left(  s\right)  \right)  - \beta
F\left(  x_{A}\right)  }\sqrt{\det\left(  g\left(  x\left(  s\right)  \right)
\right)  }}{\int_{\Sigma\left(  s\right)  }e^{-\left(  \beta F\left(
\mathbf{x}^{\left(  s\right)  }\left(  \mathbf{u}\right)  \right)  -\beta
F\left(  \mathbf{x}\left(  s\right)  \right)  \right)  }\sqrt{\det\left(
\overline{g}^{\left(  s\right)  }\left(  \mathbf{u}\right)  \right)
}d\mathbf{u}}ds\right)  \left(  \int_{V_{A}\left(  0\right)  }e^{-\left(
\beta F\left(  x\right)  -\beta F\left(  x_{A}\right)  \right)  }%
d\mathbf{x}\right)  .
\end{equation}
In the spirit of the stationary phase approximation, it is assumed that main
contribution to the (outer) integral in the first term is dominated by
contributions near the peak of the free energy $F(x(0))$ so that
\begin{equation}
\label{T1A}T^{\left(  2\right)  }\left(  x_{A}\right)  \simeq\frac{I^{\left(
\text{path}\right)  }I^{V_{A}\left(  0\right)  }}{I^{\Sigma\left(  0\right)
}}%
\end{equation}
where%
\begin{align}
I^{\left(  \text{path}\right)  }  &  =\int_{0}^{s_{A}}e^{\beta F\left(
x\left(  s\right)  \right)  -\beta F\left(  x_{A}\right)  }\sqrt{\det\left(
g\left(  x\left(  s\right)  \right)  \right)  }ds\\
I^{V_{A}\left(  0\right)  }  &  =\int_{V_{A}\left(  0\right)  }e^{-\left(
\beta F\left(  x\right)  -\beta F\left(  x_{A}\right)  \right)  }%
d\mathbf{x}\nonumber\\
I^{\Sigma\left(  0\right)  }  &  =\int_{\Sigma\left(  0\right)  }e^{-\left(
\beta F\left(  x^{(0)}\left(  \mathbf{u}\right)  \right)  -\beta F\left(
x\left(  0\right)  \right)  \right)  }\sqrt{\det\left(  \overline{g}^{\left(
0\right)  }\left(  u\right)  \right)  }d\mathbf{u}.\nonumber
\end{align}
This will be done via a stationary phase approximation.

\subsection{The plane perpendicular to the path, $I^{\Sigma\left(  s\right)
}$}

We want to approximate the integral%
\begin{equation}
I^{\Sigma\left(  s\right)  }=\int_{\Sigma\left(  s\right)  }e^{-\left(  \beta
F\left(  x\left(  \mathbf{u}^{\left(  s\right)  }\right)  \right)  -\beta
F\left(  x\left(  s\right)  \right)  \right)  }\sqrt{\det\left(  \overline
{g}\left(  \mathbf{u}^{\left(  s\right)  }\right)  \right)  }d\mathbf{u}%
^{\left(  s\right)  }.
\end{equation}
The direction of the path is
\begin{equation}
t^{I}\equiv\frac{dx^{I}}{ds}=\frac{g^{IJ}F_{J}}{\sqrt{F_{K}g^{KL}F_{L}}}%
\end{equation}
and the perpendicular vectors satisfy $0=n_{I}^{\left(  \alpha\right)  }%
t^{I}=\frac{n_{I}^{\left(  \alpha\right)  }g^{IJ}F_{J}}{\sqrt{F_{K}g^{KL}%
F_{L}}}$ so, $g$ is assumed to be positive-definite, this means that the the
derivative of the energy in the plane spanned by the vectors $n_{I}^{\left(
\alpha\right)  }g^{IJ}\equiv n^{\left(  \alpha\right)  I}$ has zero
derivative. So, using these vectors as a basis for the plane perpendicular to
the path,
\begin{equation}
x^{I}\left(  \mathbf{u}^{\left(  s\right)  }\right)  =x^{I}\left(
\mathbf{u}^{\left(  s\right)  }=0\right)  +\sum_{\alpha=1}^{D-1}u^{\left(
s\alpha\right)  }n^{\left(  \alpha\right)  I}%
\end{equation}
to expand the energy in this plane gives
\begin{align}
\beta F\left(  x\left(  \mathbf{u}^{\left(  s\right)  }\right)  \right)   &
=\beta F\left(  x\left(  s\right)  \right)  +\frac{1}{2}\sum_{\alpha,\gamma
=1}^{D-1}\beta F_{IJ}\left(  x\left(  s\right)  \right)  n_{\left(
\alpha\right)  }^{I}n_{\left(  \gamma\right)  }^{J}u_{\left(  \alpha\right)
}^{\left(  s\right)  }u_{\left(  \gamma\right)  }^{\left(  s\right)  }+...\\
&  \equiv\beta F\left(  x\left(  s\right)  \right)  +\frac{1}{2}\sum
_{\alpha,\gamma=1}^{D-1}\beta F_{\left(  \alpha\gamma\right)  }\left(
x\left(  s\right)  \right)  u_{\left(  \alpha\right)  }^{\left(  s\right)
}u_{\left(  \gamma\right)  }^{\left(  s\right)  }+...\nonumber
\end{align}
Now, choose the basis so that the matrix $\overline{F}_{\left(  \alpha
\gamma\right)  }$ is diagonal, $\overline{F}_{\left(  \alpha\gamma\right)
}=\delta_{\left(  \alpha\gamma\right)  }\lambda_{\left(  \alpha\right)  }$ so
that
\begin{equation}
\beta F\left(  x\left(  \mathbf{u}^{\left(  s\right)  }\right)  \right)
=\beta F\left(  x\left(  s\right)  \right)  +\frac{1}{2}\sum_{\alpha=1}%
^{D-1}\lambda_{\left(  \alpha\right)  }u_{\left(  \alpha\right)  }^{\left(
s\right)  2}+...
\end{equation}
Note, incidentally, that this does not mean that $g^{IK}\overline{F}%
_{KJ}n_{\left(  \alpha\right)  }^{J}=\lambda_{\left(  \alpha\right)
}n_{\left(  \alpha\right)  }^{I}$ because $g^{IK}\overline{F}_{KJ}n_{\left(
\alpha\right)  }^{J}$ could have a non-zero component in the direction of $t$
which is lost when contracting with $n_{\left(  \alpha\right)  }^{I}$. We are
therefore assuming that $g^{IK}\overline{F}_{IJ}n_{\left(  \alpha\right)
}^{J}=\lambda_{\left(  \alpha\right)  }n_{\left(  \alpha\right)  }%
^{I}+v_{\left(  \alpha\right)  }t^{I}$ for some $v_{\left(  \alpha\right)  }$.
So, the Gaussian approximation gives%
\begin{equation}
I^{\Sigma\left(  s\right)  }\simeq\sqrt{\det\left(  \overline{g}\left(
\mathbf{u}^{\left(  s\right)  }=0\right)  \right)  }\frac{\left(  2\pi\right)
^{\frac{n-1}{2}}}{\sqrt{\lambda_{\left(  1\right)  }...\lambda_{\left(
n-1\right)  }}}%
\end{equation}
Now
\begin{equation}
\lambda_{\left(  1\right)  }...\lambda_{\left(  n-1\right)  }=\frac
{\det\left(  Qg^{-1}FQ+Pg^{-1}FP\right)  }{tFt}%
\end{equation}
since%
\begin{equation}
Qg^{-1}FQ+Pg^{-1}FP=\sum_{\alpha\gamma}n_{\left(  \alpha\right)  }%
^{I}n_{\left(  \alpha\right)  R}g^{RI^{\prime}}F_{I^{\prime}J}n_{\left(
\gamma\right)  }^{J}n_{\left(  \gamma\right)  K}+\left(  tg^{-1}Ft\right)
t^{I}t_{K}=\sum_{\alpha}\lambda_{\left(  \alpha\right)  }n_{\left(
\alpha\right)  }^{I}n_{\left(  \alpha\right)  K}+\left(  tg^{-1}Ft\right)
t^{I}t_{K}%
\end{equation}
and this is clearly diagonal in the $\left(  t,n_{\left(  1\right)
},...,n_{\left(  n-1\right)  }\right)  $ orthonormal basis. So, using%
\begin{equation}
\det\left(  \overline{g}\left(  \mathbf{u}^{\left(  s\right)  }=0\right)
\right)  =\det_{\alpha\gamma}\left(  \frac{\partial x^{I}}{\partial u_{\left(
\alpha\right)  }}g_{IJ}\frac{\partial x^{J}}{\partial u_{\left(
\gamma\right)  }}\right)  =\det_{\alpha\gamma}\left(  n_{\left(
\alpha\right)  }^{I}g_{IJ}n_{\left(  \gamma\right)  }^{J}\right)
=\det_{\alpha\gamma}\left(  n_{\left(  \alpha\right)  }^{I}n_{J\left(
\gamma\right)  }\right)  =1
\end{equation}
gives finally%
\begin{align}
I^{\Sigma\left(  s\right)  }  &  \simeq\left(  2\pi\right)  ^{\frac{n-1}{2}%
}\sqrt{\frac{tFt}{\det\left(  Qg^{-1}FQ+Pg^{-1}FP\right)  }}\\
&  =\left(  2\pi\right)  ^{\frac{n-1}{2}}\sqrt{\frac{tFt}{\det\left(  \left(
1-tt^{T}\right)  g^{-1}F\left(  1-t^{T}t\right)  +tt^{T}g^{-1}Ft^{T}t\right)
}}\nonumber\\
&  =\left(  2\pi\right)  ^{\frac{n-1}{2}}\sqrt{\frac{t^{K}F_{KL}t^{L}}%
{\det_{IK}\left(  g^{IJ}F_{JK}-t^{I}t_{I^{\prime}}g^{I^{\prime}J}F_{JK}%
-g^{IJ}F_{JJ^{\prime}}t^{J^{\prime}}t_{K}+2t^{I}t_{K}\left(  t^{J}%
F_{JJ^{\prime}}t^{J^{\prime}}\right)  \right)  }}\nonumber
\end{align}

At the critical point, $t_{I}$ is the unstable eigenvector for the dynamical
matrix, so $\left(  tFt\right)  =\lambda^{\left(  -\right)  }$ and the vectors
$n_{\left(  \alpha\right)  }$ are all orthonormal eigenvectors of the
dynamical matrix so $Qg^{-1}FP^{T}=Pg^{-1}FQ^{T}=0,$e.g.
\begin{align}
Qg^{-1}FP  &  =n_{\left(  \alpha\right)  I}^{I}n_{\left(  \alpha\right)
J}g^{JJ^{\prime}}F_{J^{\prime}K}t^{K}t_{R}\\
&  =n_{\left(  \alpha\right)  I}^{I}n_{\left(  \alpha\right)  J}\left(
\lambda^{\left(  -\right)  }t^{J}\right)  t_{R}\nonumber\\
&  =\lambda^{\left(  -\right)  }n_{\left(  \alpha\right)  I}^{I}\left(
n_{\left(  \alpha\right)  J}t^{J}\right)  t_{R}\nonumber\\
&  =0,\nonumber
\end{align}
and%
\begin{align}
Pg^{-1}FP  &  =t^{I}t_{J}g^{JK}F_{KL}t^{L}t_{M}=t^{I}t_{J}\left(
\lambda^{\left(  -\right)  }t^{J}\right)  t_{M}=\lambda^{\left(  -\right)
}t^{I}t_{M}\\
Qg^{-1}FQ^{T}  &  =\sum_{\alpha=1}^{n-1}\lambda^{\left(  \alpha\right)
}n_{\left(  \alpha\right)  }^{I}n_{\left(  \alpha\right)  M}\nonumber
\end{align}
implying that $Qg^{-1}FQ+Pg^{-1}FP=g^{-1}F$ and giving
\begin{equation}
I^{\Sigma\left(  0\right)  }=\left(  2\pi\right)  ^{\frac{n-1}{2}}\sqrt
{\frac{\lambda^{\left(  -\right)  }}{\det\left(  g^{-1}F\right)  }}.
\end{equation}

\subsection{The quasi-equilibrium basin, $I^{V_{A}\left(  0\right)  }$}

We begin with the integral over the entire basin of attraction,%
\begin{equation}
I^{V_{A}\left(  0\right)  }=\int_{V\left(  0\right)  }e^{-\left(  \beta
F\left(  x\right)  -\beta F\left(  x_{A}\right)  \right)  }dx\simeq
\int_{-\infty}^{\infty}e^{-\frac{1}{2}\beta F_{IJ}\delta x^{I}\delta x^{J}%
}d\delta x.
\end{equation}
Let the solution to the eigenvectors of $\beta F_{IJ}$ be $u_{I}^{\left(
a\right)  }$ and the eigenvalues $\gamma^{\left(  a\right)  }$ so that
\begin{equation}
\beta F_{IJ}u_{J}^{\left(  a\right)  }=\gamma^{\left(  a\right)  }%
u_{I}^{\left(  a\right)  }%
\end{equation}
and since the Hessian is real and symmetric, the eigenvectors are orthonormal%
\begin{equation}
u_{I}^{\left(  a\right)  }u_{I}^{\left(  b\right)  }=\delta_{ab}.
\end{equation}
One can therefore expand as
\begin{align}
\delta x^{I}  &  =c^{\left(  a\right)  }u_{I}^{\left(  a\right)
},\;c^{\left(  a\right)  }=\delta x^{I}u_{I}^{\left(  a\right)  }\\
F_{IJ}\delta x^{I}\delta x^{J}  &  =\sum_{a}\gamma^{\left(  a\right)
}c^{\left(  a\right)  2}\nonumber
\end{align}
so%
\begin{equation}
I^{V_{A}\left(  0\right)  }=\int_{-\infty}^{\infty}e^{-\frac{1}{2}\sum
_{a}\gamma^{\left(  a\right)  }c^{\left(  a\right)  2}}\underset
{1}{\underbrace{\det\left(  u^{\left(  1\right)  }...u^{\left(  n\right)
}\right)  }}d\mathbf{c=}%
%TCIMACRO{\dprod \limits_{a=1}^{n}}%
%BeginExpansion
{\displaystyle\prod\limits_{a=1}^{n}}
%EndExpansion
\sqrt{\frac{2\pi}{\gamma^{\left(  a\right)  }}}=\frac{\left(  2\pi\right)
^{n/2}}{\det F^{\prime\prime}\left(  x^{\left(  A\right)  }\right)  }%
\end{equation}

\subsection{Along the unstable direction, $I^{\left(  \text{path}\right)  }$}

We need%
\begin{align}
I^{\left(  \text{path}\right)  }  &  =\int_{0}^{s_{A}}\sqrt{\det\left(
g\left(  x\left(  s\right)  \right)  \right)  }e^{\beta F\left(  x\left(
s\right)  \right)  -\beta F\left(  x(s_{A})\right)  }ds\\
&  \simeq\sqrt{\det\left(  g\left(  0\right)  \right)  } e^{\beta F(x^{\ast})-
\beta F(x_{A})}\int_{-\infty}^{0}e^{\frac{1}{2}\left(  \frac{d^{2}}{ds^{2}%
}F\left(  x\left(  s\right)  \right)  \right)  _{s=0}s^{2}}ds\nonumber\\
&  =\sqrt{\det\left(  g\left(  0\right)  \right)  }\sqrt{\frac{\pi
}{2\left\vert \frac{d^{2}}{ds^{2}}F\left(  x\left(  s\right)  \right)
\right\vert _{s=0}}} e^{\beta F(x^{\ast})- \beta F(x_{A})}\nonumber
\end{align}
Now,%
\begin{equation}
\frac{d^{2}}{ds^{2}} \beta F\left(  x\left(  s\right)  \right)  =\frac{d}%
{ds}\left(  \frac{dx^{I}}{ds}F_{I}\left(  x\right)  \right)  =\frac{d^{2}%
x^{I}}{ds^{2}}F_{I}\left(  x\right)  +\frac{dx^{I}}{ds}\frac{dx^{J}}{ds}%
F_{IJ}\left(  x\right)
\end{equation}
and evaluating at the critical point, remembering that $F_{I}\left(  x^{\ast
}\right)  =0$, gives%
\begin{align}
\left(  \frac{d^{2}}{ds^{2}} \beta F\left(  x\left(  s\right)  \right)
\right)  _{s=0}  &  =\left(  \frac{dx^{I}}{ds}\frac{dx^{J}}{ds}\right)
_{s=0}F_{IJ}\left(  x^{\ast}\right) \\
&  =\left(  \frac{g^{IK}\beta F_{K}}{\sqrt{\beta F_{L}g^{LM}\beta F_{M}}}%
\frac{g^{JN}\beta F_{N}}{\sqrt{\beta F_{R}g^{RS}\beta F_{S}}}\right)
_{s=0}F_{IJ}\left(  x^{\ast}\right) \nonumber
\end{align}
Now, the initial value to determine the MLP\ is exactly that $x^{I}=x^{I\ast
}+\epsilon v^{\left(  -\right)  I}$ with $\epsilon$ going to zero, so%
\begin{align}
\beta F_{K}  &  \rightarrow\beta F_{K}\left(  x^{\ast}\right)  +\epsilon\beta
F_{KL}\left(  x^{\ast}\right)  v^{\left(  -\right)  L}\\
&  =\epsilon g_{KM}\left(  x^{\ast}\right)  g^{MN}\left(  x^{\ast}\right)
\beta F_{NL}\left(  x^{\ast}\right)  v^{\left(  -\right)  L}\nonumber\\
&  =\epsilon\lambda^{\left(  -\right)  }g_{KM}\left(  x^{\ast}\right)
v^{\left(  -\right)  M}\nonumber\\
&  =\epsilon\lambda^{\left(  -\right)  }v_{K}^{\left(  -\right)  }\nonumber
\end{align}
implying that
\begin{equation}
\lim_{s\rightarrow0}\frac{g^{IK}\beta F_{K}}{\sqrt{\beta F_{L}g^{LM}\beta
F_{M}}}=\lim_{\epsilon\rightarrow0}\frac{g^{IK}\epsilon\lambda^{\left(
-\right)  }v_{K}^{\left(  -\right)  }}{\sqrt{\epsilon\lambda^{\left(
-\right)  }v_{L}^{\left(  -\right)  }g^{LM}\epsilon\lambda^{\left(  -\right)
}v_{M}^{\left(  -\right)  }}}=\frac{v^{\left(  -\right)  I}}{\sqrt
{v_{L}^{\left(  -\right)  }v^{\left(  -\right)  L}}}=v^{\left(  -\right)  I}%
\end{equation}
and%
\begin{equation}
\left(  \frac{d^{2}}{ds^{2}} \beta F\left(  x\left(  s\right)  \right)
\right)  _{s=0}=v^{\left(  -\right)  I}F_{IJ}\left(  x^{\ast}\right)
v^{\left(  -\right)  J}=\lambda^{\left(  -\right)  }%
\end{equation}
giving%
\begin{equation}
I^{\left(  \text{path}\right)  }\simeq\sqrt{\det\left(  g\left(  0\right)
\right)  }\sqrt{\frac{\pi}{2\left\vert \lambda^{\left(  -\right)  }\right\vert
}} e^{\beta F(x^{\ast})- \beta F(x_{A})}%
\end{equation}

\subsection{Final result}

Substituting these three approximations into Eq.(\ref{T1A}) gives%
\begin{align}
T^{\left(  \text{Langer}\right)  }\left(  x_{A}\right)   &  \simeq\left(
\frac{\sqrt{\det\left(  g^{\ast}\right)  }\sqrt{\frac{\pi}{2\left\vert
\lambda^{\left(  -\right)  }\right\vert }}e^{\beta F\left(  x^{\ast}\right)
-\beta F\left(  x_{A}\right)  }}{\frac{\left(  2\pi\right)  ^{\frac{n-1}{2}}%
}{\sqrt{\det\left(  g^{\ast-1}F^{\ast}\right)  }}\sqrt{\left\vert
\lambda^{\left(  -\right)  }\right\vert }}\right)  \left(  \frac{\left(
2\pi\right)  ^{\frac{n}{2}}}{\sqrt{\left\vert \det F^{\prime\prime}\left(
x_{A}\right)  \right\vert }}\right) \\
&  =\frac{\pi}{\left\vert \lambda^{\left(  -\right)  }\right\vert }\frac
{\sqrt{\det\left\vert F^{^{\prime\prime}}\left(  x^{\ast}\right)  \right\vert
}}{\sqrt{\det F^{\prime\prime}\left(  x_{A}\right)  }}e^{\beta F\left(
x^{\ast}\right)  -\beta F\left(  x_{A}\right)  }\nonumber
\end{align}
as reported in the main text. Note that the matrix $g^{\ast}$ appears in the
individual terms but cancels out of the final expression.

\section{Simulations}

\label{Simulations}

Our simulations use the simple Euler-Maruyama method\cite{Kloeden}. So the equations%
\begin{equation}
\frac{dx^{I}}{dt}=-g^{IJ}\left(  x\right)  \frac{\partial\beta F\left(
x\right)  }{\partial x^{J}}+q^{I(a)}\left(  x\right)  \widehat{\xi}^{\left(
a\right)  }\left(  t\right)
\end{equation}
with the Ito interpretation is simulated as
\begin{equation}
x_{t+\tau}^{I}=x_{t}^{I}+\tau\left(  -g^{IJ}\left(  x\right)  \frac
{\partial\beta F\left(  x\right)  }{\partial x^{J}}\right)  _{x_{\tau}}%
+\sqrt{\tau}q^{I(a)}\left(  x_{t}\right)  \xi_{t}^{\left(  a\right)  }%
\end{equation}
with the numbers $\xi_{t}^{\left(  a\right)  }$ sampled from a gaussian
distribution with unit variance. If the equation is combined with the anti-Ito
interpretation, we must use%
\begin{align}
x_{t+\tau}^{I}  &  =x_{t}^{I}+\tau\left(  -g^{IJ}\left(  x\right)
\frac{\partial\beta F\left(  x\right)  }{\partial x^{J}}+\frac{\partial
}{\partial x^{J}}g^{IJ}\left(  x\right)  \right)  _{x_{\tau}}+\sqrt{\tau
}q^{I(a)}\left(  x_{t}\right)  \xi_{t}^{\left(  a\right)  }\\
&  =x_{t}^{I}+\tau\left(  -g^{IJ}\left(  x\right)  \frac{\partial\beta
F\left(  x\right)  }{\partial x^{J}}+\frac{\partial q^{I\left(  a\right)
}\left(  x\right)  }{\partial x^{J}}q^{J\left(  a\right)  }\left(  x\right)
+q^{I\left(  a\right)  }\left(  x\right)  \frac{\partial q^{J\left(  a\right)
}\left(  x\right)  }{\partial x^{J}}\right)  _{x_{\tau}}+\sqrt{\tau}%
q^{I(a)}\left(  x_{t}\right)  \xi_{t}^{\left(  a\right)  }.\nonumber
\end{align}

\end{document}